\documentclass[11pt]{article}
\usepackage[utf8]{inputenc}
\usepackage{amsfonts}
\usepackage{amsmath}
\usepackage{amssymb}
\usepackage{indentfirst}
\usepackage{graphicx}
\usepackage[colorlinks]{hyperref}
\usepackage{cite}
\usepackage{colortbl}
\usepackage{microtype}
\usepackage[normalem]{ulem}
\usepackage{xcolor}
\usepackage{cancel}
\usepackage{soul}
\usepackage{caption}

\DeclareCaptionType{equ}[][]
\numberwithin{equation}{section}
\allowdisplaybreaks

\begin{document}

\baselineskip=18pt %a la harvmac
\baselineskip 0.6cm
\begin{titlepage}
\vskip 4cm

\begin{center}
\textbf{\LARGE{Three-dimensional Hypergravity Theories and Semigroup Expansion Method}}
\par\end{center}{\LARGE \par}

\begin{center}
	\vspace{1cm}
	\textbf{Ricardo Caroca}$^{\dag,\bullet}$,
	\textbf{Patrick Concha}$^{\ast}$,
    \textbf{Javier Matulich}$^{\ddag}$,
	\textbf{Evelyn Rodríguez}$^{\ast}$,\\
	\textbf{David Tempo}$^{\star}$
	\small
	\\[5mm]
 $^{\dag}$\textit{Instituto de Ciencias Exactas y Naturales,}\\\textit{Universidad Arturo Prat, Playa Brava 3265, 1111346, Iquique, Chile.}
 \\[2mm]
 $^{\bullet}$\textit{Facultad de Ciencias, Universidad Arturo Prat,}\\\textit{Avenida Arturo Prat Chacón 2120, 1110939, Iquique, Chile.}
 \\[2mm]
    $^{\ast}$\textit{Departamento de Matemática y Física Aplicadas, }\\
	\textit{ Universidad Católica de la Santísima Concepción, }\\
\textit{ Alonso de Ribera 2850, Concepción, Chile.}
	\\[2mm]
	$^{\ddag}$\textit{Instituto de Física Teórica UAM/CSIC,}\\
	\textit{C/ Nicolás Cabrera 13-15, Universidad Autónoma de Madrid, Cantoblanco, Madrid 28049,Spain. }
	\\[2mm]
	$^{\star}$\textit{Departamento de Ciencias Matemáticas y Físicas, Facultad de Ingeniería}\\
	\textit{Universidad Católica de Temuco, Chile.}\\[5mm]
	\footnotesize
	\texttt{rcarocaster@gmail.com},
	\texttt{patrick.concha@ucsc.cl},
    \texttt{javier.matulich@csic.es},
	\texttt{erodriguez@ucsc.cl},
	\texttt{jtempo@uct.cl}
	\par\end{center}
\vskip 26pt

%\preprint{IFT-UAM/CSIC-23-44}
%\begin{abstract}
\centerline{{\bf Abstract}}
\medskip
\noindent  
In this work we present novel and known three-dimensional hypergravity theories which are  obtained by applying the powerful semigroup expansion method. We show that the expansion procedure considered here yields a consistent way of coupling different three-dimensional Chern-Simons gravity theories with massless spin-$\frac{5}{2}$ gauge fields.  First, by expanding the $\mathfrak{osp}\left(1|4\right)$ superalgebra with a particular semigroup a generalized hyper-Poincaré algebra is found. Interestingly, the hyper-Poincaré and hyper-Maxwell algebras appear as subalgebras of this generalized hypersymmetry algebra. Then, we show that the generalized hyper-Poincaré CS gravity action can be written as a sum of diverse hypergravity CS Lagrangians. We extend our study to a generalized hyper-AdS gravity theory by considering a different semigroup. Both generalized hyperalgebras are then found to be related through an Inönü-Wigner contraction which can be seen as a generalization of the existing vanishing cosmological constant limit between the hyper-AdS and hyper-Poincaré gravity theories.

%\end{abstract}
\vspace{-23cm}
\begin{flushright}
{\footnotesize IFT-UAM/CSIC-23-44}
\end{flushright}
\vspace{2cm}
\end{titlepage}\newpage {\baselineskip=12pt \tableofcontents{}}

\section{Introduction}\label{sec1}

Three-dimensional higher spin (HS) gauge theories \cite{Aragone:1983sz,Vasiliev:1986qx,Blencowe:1988gj,Bergshoeff:1989ns,Vasiliev:1995dn,Campoleoni:2010zq}
have proven to be exceptional setups for a deeper understanding of
some relevant features of higher spin gauge theories in four and higher
dimensions \cite{Fradkin:1987ks,Vasiliev:1990en,Vasiliev:2003ev}.
Indeed, despite the fact that the generic theory possesses an infinite
tower of higher spin fields, it can be simplified leading to a consistent
truncation endowed with a finite number of non-propagating though
interacting higher spin fields whose dynamics can be described in
terms of a Chern--Simons (CS) action \cite{Blencowe:1988gj,Bergshoeff:1989ns,Vasiliev:1995dn}.
In particular, these toy models can be compactly written using the
well known framelike formulation of three-dimensional Einstein gravity as a CS gauge
theory where the gauge algebra can be chosen as $\mathfrak{so}(2,2)\simeq \mathfrak{sl}(2,\mathbb{R})\oplus \mathfrak{sl}(2,\mathbb{R})$
or $\mathfrak{iso}(2,1)\simeq\mathbb{R}^{2,1}\rtimes \mathfrak{sl}(2,\mathbb{R})$, for
Anti-de Sitter (AdS$_{3}$) gravity or flat space gravity, respectively
\cite{Achucarro:1986uwr,Witten:1988hc}. Thus, these special classes
of bosonic higher spin gauge theory in three spacetime dimensions can be built by simply
enlarging the corresponding $\mathfrak{sl}(2,\mathbb{R})$ (sub)algebra(s) to
a HS algebra that includes it \cite{Aragone:1983sz,Vasiliev:1986qx,Blencowe:1988gj,Bergshoeff:1989ns,Vasiliev:1995dn};
which can be finite-dimensional ($\mathfrak{sl}(N,\mathbb{R})$) or infinite-dimensional
($hs(\lambda$)). 

The locally supersymmetric extensions of these HS theories also admit
a consistent formulation by means of a CS theory for suitable graded
versions (see e.g., \cite{Henneaux:2012ny}) of the corresponding
HS algebra by following the same line of reasoning used in the supergravity
extensions of General Relativity in three spacetime dimensions for either negative \cite{Achucarro:1987vz}
or vanishing cosmological constant \cite{Achucarro:1989gm,Howe:1995zm}.
One prominent example among these supersymmetric extensions is the
AdS$_{3}$ hypergravity that consists in a HS field theory containing
fields of spins 2, 4 and 5/2 being invariant under the action of higher
spin fermionic symmetry transformations (hypersymmetry) with spin-3
parameters. Indeed, this theory is described by a CS theory based
on the gauge $\mathfrak{osp}(1|4)\oplus \mathfrak{osp}(1|4)$ superalgebra, where $\mathfrak{sl}(2,\mathbb{R})$ is principally embedded in $\mathfrak{sp}(4)$ \cite{Zinoviev:2014sza,Henneaux:2015tar,Henneaux:2015ywa}.
Noticeably, in the vanishing cosmological constant limit the spin-4
field completely decoupled such that the HS theory reduces to the
one obtained by Aragone and Deser in \cite{Aragone:1983sz} (see also
\cite{Fuentealba:2015jma,Fuentealba:2015wza}).

In a previous work \cite{Caroca:2021bjo} a consistent way of coupling three-dimensional Maxwell-CS gravity theory with a spin-$\frac{5}{2}$ gauge field was explored. The Maxwell algebra has been first introduced in \cite{Bacry:1970ye,Bacry:1970du,Schrader:1972zd} to describe a Minkowski spacetime in presence of a electromagnetic background field \cite{Gomis:2017cmt}. Subsequently, diverse (super)gravity models based on the Maxwell symmetries have been studied in \cite{Cangemi:1992ri,Duval:2008tr,Gomis:2009dm,Bonanos:2010fw,Durka:2011nf,deAzcarraga:2012qj,Salgado:2014jka,Hoseinzadeh:2014bla,Ravera:2018vra,Aviles:2018jzw,Concha:2018zeb,Concha:2018jxx,Concha:2019icz,Salgado-Rebolledo:2019kft,Chernyavsky:2020fqs,Adami:2020xkm,Kibaroglu:2020tbr,Concha:2021jnn,Cebecioglu:2022iyq} with diverse purposes. At the bosonic level, the Maxwell algebra has been useful to accommodate a cosmological constant in four spacetime dimensions \cite{deAzcarraga:2010sw}. On the other hand, its generalizations known as generalized Poincaré algebras\footnote{The generalized Poincaré algebra is also denoted as $\mathfrak{B}_{k}$ in the literature and has been studied in diverse contexts \cite{Concha:2014zsa,Concha:2016hbt,Caroca:2017izc,Concha:2020sjt,Izaurieta:2021hfx}.} have been used to extend  standard general relativity from CS and Born-Infeld gravity \cite{Edelstein:2006se,Izaurieta:2009hz,Concha:2013uhq,Concha:2014vka}. The asymptotic symmetry of the Maxwell CS gravity theory has then been presented in \cite{Concha:2018zeb} showing the consequences of the extra Maxwellian gauge field at the level of the dynamics boundary.

There are several procedures to obtain new algebras such as contractions \cite{Inonu:1953sp, doi:10.1063/1.1724208, doi:10.1063/1.530905}, deformations \cite{Gerstenhaber:1963zz,bams/1183527432,Fialowski_2006}, extensions \cite{deAzcarraga:2004zj, Hoseinzadeh:2014bla}, double extensions \cite{Medina1985,Figueroa-OFarrill:1995opp,Matulich:2019cdo}, and expansions \cite{Hatsuda:2001pp,deAzcarraga:2002xi,Izaurieta:2006zz,deAzcarraga:2007et}, among others. In the context of expansion of algebras, the semigroup expansion method has been particularly fruitful, which is based on combining the structure constants of the Lie algebra with the inner product of an abelian semigroup $S$, in order to define the Lie brackets of a new $S$-expanded Lie algebra \cite{Izaurieta:2006zz}. This $S$-expansion method allows to obtain various expanded algebras for different semigroups, along with their invariant tensors, thus making  this scheme particularly useful to construct invariant actions under new symmetries by means of Chern-Simons or transgression forms. 

The $S$-expansion method offers a large and diverse catalogue of both known and novel expanded algebras and superalgebras, which have allowed to construct and study an equally large amount of action principles invariant under interesting symmetries \cite{Izaurieta:2006aj,deAzcarraga:2007et,deAzcarraga:2012zv,Concha:2014tca,Concha:2015tla,Caroca:2018obf,Bergshoeff:2019ctr,deAzcarraga:2019mdn,Penafiel:2019czp,Gomis:2019nih,Fontanella:2020eje,Concha:2020ebl,Concha:2020eam,Kasikci:2021atn,Gomis:2022spp,Caroca:2022byi}. A question that arises is whether it is possible to extend the already known hypergravity theories to other hyperalgebras, for example those resulting by expanding the $\mathfrak{osp}(4|1)$ superalgebra. In this work, we show that the $S$-expansion method offers us a straightforward procedure to obtain different CS gravity theories consistently coupled with spin-$\frac{5}{2}$ gauge fields. We not only obtain already known hypergravities CS models but also novel ones, which contain the hyper-Poincaré and hyper-Maxwell as particular subcases. The extension to generalized hyper-AdS gravity theories by considering a different semigroup family is also studied. Finally, both generalized hyperalgebras are shown to be related through an Inönü-Wigner (IW) contraction which can be seen as a generalization of the existing vanishing cosmological constant limit between the hyper-AdS and hyper-Poincaré gravity theories.

Thus, our results extend the expansion and contraction relations existing in the spin-2 case to the context of higher-spin algebras.
\begin{equ}[!h]
\begin{equation*}
\begin{tabular}{ccc}
\cline{3-3}

&  & \multicolumn{1}{|c|}{$\mathfrak{so}\left(2,2\right)$} \\ \cline{3-3}
& $\nearrow _{S_{\mathcal{M}}^{\left( 1\right) }}$ &  \\ \cline{1-1}

\multicolumn{1}{|c}{$\mathfrak{so}\left(2,1\right)$}& \multicolumn{1}{|c}{} & $\downarrow $ \
$\ell \rightarrow \infty $ \\ \cline{1-1}
& $\searrow ^{S_{E}^{\left( 1\right) }}$ &  \\ \cline{3-3}

&  & \multicolumn{1}{|c|}{$\mathfrak{iso}\left(2,1\right)$} \\ \cline{3-3}
\end{tabular}%
\overset{%
\begin{array}{c}
\text{{\small Generalization}}%
\end{array}%
}{\longrightarrow }%
\begin{tabular}{ccc}
\cline{3-3}

&  & \multicolumn{1}{|c|}{Generalized AdS} \\
 \cline{3-3}
& $\nearrow _{S_{\mathcal{M}}^{\left( N\right) }}$ &  \\ \cline{1-1}

\multicolumn{1}{|c}{$\mathfrak{so}\left(2,1\right)$}& \multicolumn{1}{|c}{} & $\downarrow$ \ IW \\ \cline{1-1}
& $\searrow ^{S_{E}^{\left( N\right) }}$ &  \\ \cline{3-3}

&  & \multicolumn{1}{|c|}{Generalized Poincaré} \\
\cline{3-3}
\end{tabular} 
\end{equation*}
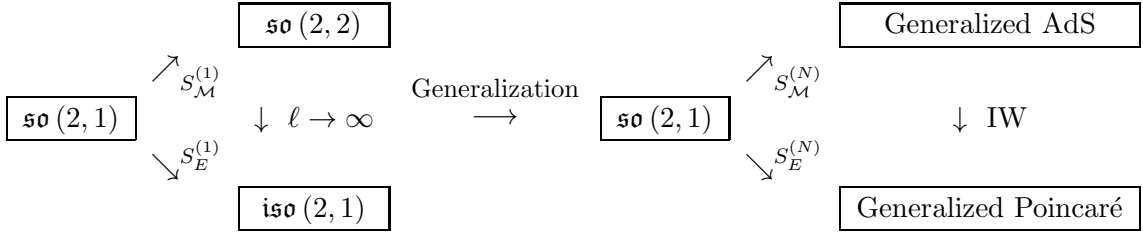
\caption{S-expansions of the $\mathfrak{so}\left(2,1\right)$ algebra carried out in terms of suitable semigroups S.}
\label{diagram1}
\end{equ}
Interestingly, diagram \eqref{diagram1} summarizing the different S-expansions of the $\mathfrak{so}\left(2,1\right)$ algebra \cite{Caroca:2017onr} can be naturally extended to yield two inequivalent families of hyper-algebras starting from the $\mathfrak{osp}\left(1|4\right)$ superalgebra\footnote{ The explicit multiplication law of the semigroups appearing in the diagram can be found in section \ref{sec3} and \ref{sec4}.}. Diagram \eqref{diagram} summarizes the novel relations obtained for the different hyper-algebras.

The paper is organized as follows. In section \ref{sec2} we present a brief review of the $osp(4|1)$ superalgebra in a convenient basis for our purposes. Then, a generic expansions of this algebra is performed in terms of an arbitrary semigroup $S$ along with its expanded Chern-Simons action in $2+1$ dimensions. Section \ref{sec3} is devoted to obtain known and novel hypersymmetric extensions of the Poincaré and Maxwell algebra by means of the $S$-expansion method. The hypergravity theories invariant under each symmetry are also constructed. In section \ref{sec4} different semigroups allow to construct novel generalized hyper-AdS algebras and their corresponding hypergravity theories. Section  \ref{sec5} is devoted to some discussion and future developments.

\section{The $\mathfrak{osp}\left(1|4\right)$ Chern-Simons action and semigroup expansion}\label{sec2}
In this section, we first review the $\mathfrak{osp}\left(1|4\right)$ superalgebra and the corresponding CS action based on it. We then present a generic expanded $\mathfrak{osp}\left(1|4\right)$ superalgebra for an arbitrary semigroup $S$, along with its expanded CS action. As we shall see, different semigroups will allow us to obtain known and novel hyper-algebras as well as the associated three-dimensional hypergravity actions.

\subsection{$\mathfrak{osp}\left(1|4\right)$ superalgebra and Chern-Simons action} \label{sec2.1}
The $\mathfrak{osp}\left( 1|4\right) $ superalgebra with $\mathfrak{sl}\left( 2,\mathbb{R}\right) $ principal embedded in $\mathfrak{sp}\left(4\right)$  
 is spanned by the set of generators $\left\{
T_{a},T_{abc},\mathcal{G}_{\alpha a}\right\} $ which satisfy the following non-vanishing
(anti-)commutation relations
\begin{eqnarray}
\left[ T_{a},T_{b}\right] &=&\epsilon _{~ab}^{m}T_{m}\,,  \notag \\
\left[ T_{a},T_{bcd}\right] &=&3\epsilon _{~a\left( b\right. }^{m}T_{\left.
cd\right) m}\,,  \notag \\
\left[ T_{a},\mathcal{G}_{\alpha b}\right] &=&\frac{1}{2}\left( \Gamma
_{a}\right) _{\text{ }\alpha }^{\beta }\mathcal{G}_{\beta b}+\epsilon _{abc}%
\mathcal{G}_{\beta }^{~c}\,,  \notag \\
\left[ T_{abc},T_{mnk}\right] &=&-6\left( \eta _{\left( ab\right. }\epsilon
_{~\left. c\right) \left( m\right. }^{l}\eta _{\left. nk\right) }+5\epsilon
_{~\left( m\right. |\left( a\right. }^{l}\delta _{~b}^{d}\eta _{\left.
c\right) |n}\eta _{\left. k\right) d}\right) T_{l} \notag  \\
&&+2\left( 5\epsilon _{~\left( m\right. |\left( a\right. }^{l}\delta
_{~b}^{d}T_{\left. c\right) l|n}\eta _{\left. k\right) d}-\epsilon _{~\left(
m\right. |\left( a\right. }^{l}\eta _{\left. bc\right) }T_{|\left. nk\right)
l}-\epsilon _{~\left( m\right. \left( a\right. |}^{l}T_{\left. bc\right)
l}\eta _{|\left. nk\right) }\right) \,,  \notag \\
\left[ T_{abc},\mathcal{G}_{\alpha d}\right] &=&\left( \delta _{~d}^{k}\eta
_{\left( ab\right. |}-5\eta _{d\left( a\right. }\delta _{~b|}^{k}\right)
\left( \Gamma _{|\left. c\right) }\right) _{~\alpha }^{\beta }\mathcal{G}%
_{\beta k}+\eta _{\left( ab\right. |}\left( \Gamma _{d}\right) _{~\alpha
}^{\beta }\mathcal{G}_{\beta |\left. c\right) }\,,  \notag \\
\left\{ \mathcal{G}_{\alpha a},\mathcal{G}_{\beta b}\right\} &=&\left(
T_{abc}-\frac{4}{3}\eta _{ab}T_{c}\right) \left( C\Gamma ^{c}\right)
_{\alpha \beta }+\frac{5}{3}\epsilon _{abc}C_{\alpha \beta }T^{c}+\frac{2}{3}%
T_{\left( a\right. |}\left( C\Gamma _{|\left. b\right) }\right) _{\alpha
\beta }\,.  \label{osp14}
\end{eqnarray}%
where $a,b,\dots =0,1,2$ are Lorentz indices lowered and raised with the
off-diagonal Minkowski metric whose nonvanishing components are given by $\eta_{01} = \eta_{10}=\eta_{22}=1$, and $\epsilon_{abc}$ is the three-dimensional Levi-Civita tensor. The $\Gamma$'s denote the gamma matrices on three spacetime dimensions and $C$ is the charge conjugation matrix satisfying $C^T=-C$ and $C\Gamma^{a}=\left(C\Gamma^{a}\right)^{T}$. Here $T_{a}$ stand for the spin-2
generators which span the $%
\mathfrak{sl}\left( 2,\mathbb{R}\right) $ subalgebra. On the other hand,  $T_{abc}$ correspond to spin-4 generators, while the fermionic $\mathcal{G}_{\alpha a}$ generators are spin$-\frac{5}{2}$ generators, which transform in a spin-$3/2$ irreducible representation of the Lorentz group. The spin-4 generators are
traceless and totally symmetric satisfying $\eta ^{ab}T_{abc}=0$ while $\mathcal{G}_{\alpha a}$ are $\Gamma $-traceless vector-spinor
generators satisfying $\left( \Gamma ^{a}\right) _{\text{ }\alpha }^{\beta }%
\mathcal{G}_{\beta a}=\Gamma ^{a}\mathcal{G}_{a}=0$.  One can note that the subalgebra spanned by the set $\{T_a,T_{abc}\}$ defines a $\mathfrak{sp}(4)$ algebra. 

Along this work, the convention adopted for the symmetrization of a pair of indices $a$, $b$ denoted as $\left(ab\right)$ is performed with a normalization factor, i.e.,
\begin{equation}
    T_{\left(ab\right)}=\frac{1}{2!}\left(T_{ab}+T_{ba}\right)\,.
\end{equation}
The $\mathfrak{osp}\left(1|4\right)$ superalgebra admits the following invariant bilinear trace
\begin{eqnarray}
\left\langle T_{a}T_{b}\right\rangle &=&\mu\eta _{ab}\,,  \notag \\
\left\langle T_{abc}T_{mnk}\right\rangle &=&2\mu\left(5\eta _{m\left( a\right.
}\eta _{b|n}\eta _{|\left. c\right) k}-3\eta _{\left( ab\right. }\eta
_{|\left. c\right) \left( m\right. }\eta _{\left. nk\right) }\right) \,, \notag \\
\left\langle \mathcal{G}_{\alpha a}\mathcal{G}_{\beta b}\right\rangle &=&%
\mu\left(\frac{4}{3}C_{\alpha \beta }\eta _{ab}-\frac{2}{3}\epsilon _{abc}\left(
C\Gamma ^{c}\right) _{\alpha \beta }\right)\,,  \label{IT1}
\end{eqnarray}
where $\mu$ is an arbitrary constant. The gauge connection one-form $A=A^{A}T_{A}$ taking value on the $\mathfrak{osp}\left(1|4\right)$ superalgebra reads
\begin{eqnarray}
   A&=&W^{a}T_{a}+W^{abc}T_{abc}+\bar{\Psi}^{a}\mathcal{G}_{a}\,, \label{1F1} 
\end{eqnarray}
where $W^{a}$ is the spin-connection, $W^{abc}$ corresponds to the spin-4 gauge fields and $\Psi^{a}$ is a Majorana spin-$\frac{5}{2}$ gauge field. The corresponding curvature two-form $F=dA+\frac{1}{2}\left[A,A\right]$ is given by
\begin{eqnarray}
    F&=&F^{a}T_{a}+F^{abc}T_{abc}+\nabla\bar{\Psi}^{a}\mathcal{G}_{a}\,, \label{2F1}
\end{eqnarray}
where
\begin{align}
    F^{a}&=dW^{a}+\frac{1}{2}\epsilon^{a}_{\ bc}W^{b}W^{c}+15\epsilon^{a}_{bc}W^{bmn}W^{c}_{\ mn}-\frac{3}{2}i\bar{\Psi}_{b}\Gamma^{a}\Psi^{b}\,, \notag \\
    F^{abc}&=dW^{abc}-5\epsilon^{(a}_{\ mn}W^{m k|b}W^{c)n}_{\ \ \ k}+3\epsilon^{(a}_{\ mn}W^{m}W^{n|bc)}+\frac{i}{2}\bar{\Psi}^{(a}\Gamma^{|b}\Psi^{c)}\,, \notag \\
    \nabla\Psi^{a}&=d\Psi^{a}+\frac{3}{2}W^{b}\Gamma_{b}\Psi^{a}-W_{b}\Gamma^{a}\Psi^{b}-5W^{bca}\Gamma_{b}\Psi_{c}\,. \label{2F1a}
\end{align}
Let us note that the ferminonic gauge fields, as the fermionic generators, are assumed to be $\Gamma$-traceless, i.e, $\Gamma^{a}\Psi_{a}=0$. A CS action for the $\mathfrak{osp}\left(1|4\right)$ superalgebra is obtained by replacing the gauge connection one-form \eqref{1F1} and the non-vanishing components of the invariant tensor \eqref{IT1} in the general expression for the CS action:
\begin{eqnarray}
    I_{CS}&=&\frac{k}{4\pi}\int\langle AdA+\frac{2}{3}A^{3} \rangle \,.\label{CS}
\end{eqnarray}
Then, the $\mathfrak{osp}\left(1|4\right)$ CS action is given, up to boundary terms, by
\begin{align}
I_{\mathfrak{osp}\left(1|4\right)}&=\frac{k}{4\pi}\int \mathcal{L}(W) +10\left(dW^{abc}-\frac{10}{3}\epsilon^{a}_{\ mn}W^{mkb}W^{cn}_{\ \ k}+3\epsilon^{a}_{\ mn}W^{m}W^{nbc}\right)W_{abc}+2i\bar{\Psi}_{a}\nabla\Psi^{a}\,, \label{ospCS}
\end{align}
where we have set $\mu=1$ and $\mathcal{L}\left(W\right)$ is the usual Lorentz-CS form \cite{Witten:1988hc} given by
\begin{align}
    \mathcal{L}(W)=  W_{a}dW^{a}+ \frac{1}{3}\epsilon _{~bc}^{a}W_{a}W^{b}W^{c} \,.  \label{LorCS}
\end{align}
The field equations of the $\mathfrak{osp}\left(1|4\right)$ CS theory are given by the vanishing of the curvature two-forms \eqref{2F1a}.

Let us note that the hyper-Poincaré algebra in presence of spin-4 generators can be obtained by applying an IW contraction to the $\mathfrak{osp}\left(1|4\right)\otimes\mathfrak{sp}\left(4\right)$ superalgebra \cite{Fuentealba:2015wza}. More recently, inequivalent hyper-Maxwell structures have been introduced by considering the IW contractions of $\mathfrak{osp}\left(1|4\right)\otimes\mathfrak{osp}\left(1|4\right)\otimes\mathfrak{sp}\left(4\right)$ and $\mathfrak{osp}\left(1|4\right)\otimes\mathfrak{sp}\left(4\right)\otimes\mathfrak{sp}\left(4\right)$ superalgebras \cite{Caroca:2021bjo}. As we shall see, those hyperalgebras along novel ones can be derived directly from one $\mathfrak{osp}\left(1|4\right)$ superalgebra considering the Lie algebra expansion method based on semigroups \cite{Izaurieta:2006zz}. Such procedure will be useful to derive the corresponding invariant trace which will be necessary for the construction of known and new CS hypergravity actions.

\subsection{S-expanded $\mathfrak{osp}\left(1|4\right)$ superalgebra} \label{sec2.2}
 Before approaching the explicit derivation of different hyperalgebras, we present the generic expanded $\mathfrak{osp}\left(1|4\right)$ superalgebra for an arbitrary semigroup $S$.
Let us first consider a subspace decomposition of the original $\mathfrak{osp}\left(1|4\right)$ superalgebra as
\begin{eqnarray}
    \mathfrak{osp}\left(1|4\right)=V_0\oplus V_1\,, \notag
\end{eqnarray}
where $V_0$ corresponds to a bosonic subspace, while $V_1$ is spanned by fermionic generators. Such decomposition satisfies a graded Lie algebra,
\begin{align}
    \left[V_0,V_0\right]&\subset V_0\,, &\left[V_0,V_1\right]&\subset V_1\,, &\left[V_1,V_1\right]&\subset V_0\,. \label{SD}
\end{align}
Let $S=\{\lambda_i\}$ be the relevant semigroup whose elements satisfy a given multiplication law. An expanded $\mathfrak{osp}\left(1|4\right)$ superalgebra is obtained by considering the direct product $S\times \mathfrak{osp}\left(1|4\right)$. The expanded generators are related to the original ones through the semigroup elements as
\begin{align}
    T^{(i)}_{a}&=\lambda_i T_{a}\,, & T^{(i)}_{abc}&=\lambda_i T_{abc}\,, & \mathcal{G}^{(i)}_{\alpha a}&=\lambda_i\mathcal{G}_{\alpha a}\,. \label{EG1}
\end{align}
The explicit (anti-)commutation relations of the expanded superalgebra are obtained considering the multiplication law of a given semigroup $S$ along with the original commutation relations of the $\mathfrak{osp}\left(1|4\right)$ superalgebra:
\begin{align}
\left[ T^{(i)}_{a},T^{(j)}_{b}\right] &=K^{\ p}_{ij}\epsilon _{~ab}^{m}T^{(p)}_{m}\,,  \notag \\
\left[ T^{(i)}_{a},T^{(j)}_{bcd}\right] &=3K^{\ p}_{ij}\epsilon _{~a\left( b\right. }^{m}T^{(p)}_{\left.
cd\right) m}\,,  \notag \\
\left[ T^{(i)}_{a},\mathcal{G}^{(j)}_{\alpha b}\right] &=\frac{1}{2}K^{\ p}_{ij}\left( \Gamma
_{a}\right) _{\ \alpha }^{\beta }\mathcal{G}^{(p)}_{\beta b}+K^{\ p}_{ij}\epsilon _{abc}%
\mathcal{G}^{~c\,(p)}_{\beta }\,,  \notag \\
\left[ T^{(i)}_{abc},T^{(j)}_{mnk}\right] &=-6K^{\ p}_{ij}\left( \eta _{\left( ab\right. }\epsilon
_{~\left. c\right) \left( m\right. }^{l}\eta _{\left. nk\right) }+5\epsilon
_{~\left( m\right. |\left( a\right. }^{l}\delta _{~b}^{d}\eta _{\left.
c\right) |n}\eta _{\left. k\right) d}\right) T^{(p)}_{l} \notag  \\
&+2K^{\ p}_{ij}\left( 5\epsilon _{~\left( m\right. |\left( a\right. }^{l}\delta
_{~b}^{d}T^{(p)}_{\left. c\right) l|n}\eta _{\left. k\right) d}-\epsilon _{~\left(
m\right. |\left( a\right. }^{l}\eta _{\left. bc\right) }T^{(p)}_{|\left. nk\right)
l}-\epsilon _{~\left( m\right. \left( a\right. |}^{l}T^{(p)}_{\left. bc\right)
l}\eta _{|\left. nk\right) }\right) \,,  \notag \\
\left[ T^{(i)}_{abc},\mathcal{G}^{(j)}_{\alpha d}\right] &=K^{\ p}_{ij}\left( \delta _{~d}^{k}\eta
_{\left( ab\right. |}-5\eta _{d\left( a\right. }\delta _{~b|}^{k}\right)
\left( \Gamma _{|\left. c\right) }\right) _{~\alpha }^{\beta }\mathcal{G}^{(p)}%
_{\beta k}+K^{\ p}_{ij}\eta _{\left( ab\right. |}\left( \Gamma _{d}\right) _{~\alpha
}^{\beta }\mathcal{G}^{(p)}_{\beta |\left. c\right) }\,,  \notag \\
\left\{ \mathcal{G}^{(i)}_{\alpha a},\mathcal{G}^{(j)}_{\beta b}\right\} &=K^{\ p}_{ij}\left(
T^{(p)}_{abc}-\frac{4}{3}\eta _{ab}T^{(p)}_{c}\right) \left( C\Gamma ^{c}\right)
_{\alpha \beta }+\frac{5}{3}K^{\ p}_{ij}\epsilon _{abc}C_{\alpha \beta }T^{c\,(p)}+\frac{2}{3}K^{\ p}_{ij}%
T^{(p)}_{\left( a\right. |}\left( C\Gamma _{|\left. b\right) }\right) _{\alpha
\beta }\,,  \label{exposp}
\end{align}%
where $K_{ij}^{\ p}$ is the so-called 2-selector defined by
\begin{equation}
K_{ij}^{\ p}=\left\{
\begin{array}{lcl}
1\,,\,\,\, & \mathrm{when}\,\,\,\,\lambda_i\lambda_{j} =\lambda_{p}\,, &
\\
0\,, & \mathrm{otherwise}\,.\ \ \ \ \ \ \ \ \ &
\end{array}%
\right.  \label{selector}
\end{equation}%
A smaller algebra can be extracted by considering a zero element $0_S\in S$. In such case a $0_S$-reduced (super)algebra of the expanded one is obtained by imposing the condition $0_ST_A=0$. Alternatively, a resonant subalgebra can be obtained by considering a semigroup decomposition $S=S_0\cup S_1$ which has to satisfy the same algebraic structure than the subspaces \eqref{SD}, namely
\begin{align}
    S_0\cdot S_0&\subset S_0\,, &S_0\cdot S_1 &\subset S_1\,, &S_1 \cdot S_1&\subset S_0\,. \label{SeD}
\end{align}
In the next sections, we will see how different semigroups, with different number of elements and satisfying a given multiplication law, lead to two inequivalent families of hyperalgebras which will be related by an Inönü-Wigner contraction. The CS hypergravity actions invariant under the aforesaid hyperalgebras will be also constructed.
%%%%%%%%%%%%%%%%%%%%%%%%%%%%%%%%%%%%%%%%%%%%%%%%%%%%%%%%%%%%%%%%%%%%%%%%%%%%%%%%%%%%%%%%%%%%%%%%%%%

\section{Generalized hyper-Poincaré gravity theory}\label{sec3}
In this section, we show that different semigroups allow us to obtain known and new hyperalgebras. It is important to mention that the choice of the semigroup is not arbitrary but is inherited from the bosonic behavior. As it was shown in \cite{Caroca:2017onr}, the Poincaré and Maxwell algebras are particular cases of the generalized Poincaré algebra \cite{Edelstein:2006se,Izaurieta:2009hz,Concha:2014zsa},  which in turn corresponds to an expansion of the $\mathfrak{so}\left(2,1\right)$ algebra with $S_{E}^{\left(N\right)}$ as the relevant semigroup (see diagram \eqref{diagram1}).  Here, using the same semigroup\footnote{Due to the presence of fermionic generator, the relevant semigroup is $S_{E}^{\left(2N\right)}$ instead of $S_{E}^{(N)}$.}, we derive the respective generalized hyper-Poincaré algebra by expanding the $\mathfrak{osp}\left(1|4\right)$ superalgebra. A we will see, for each value of $N$ a different hyperalgebra  appears. We then present the explicit construction of the respective hypergravity CS action invariant under the aforementioned hyperalgebras. 

\subsection{Hyper-Poincaré gravity} 
Here we show that the hyper-Poincaré CS theory \cite{Fuentealba:2015jma,Fuentealba:2015wza} can alternatively be obtained by expanding the $\mathfrak{osp}\left(1|4\right)$ CS theory. To this end, we consider $S_{E}^{\left(2\right)}=\{\lambda_0,\lambda_1,\lambda_2,\lambda_3\}$ as the relevant semigroup whose elements satisfy the following multiplication law:
\begin{equation}
\lambda _{\alpha }\lambda _{\beta }=\left\{ 
\begin{array}{lcl}
\lambda _{\alpha +\beta }\,\,\,\, & \mathrm{if}\,\,\,\,\alpha +\beta \leq
3\,, &  \\ 
\lambda _{3}\,\,\, & \mathrm{if}\,\,\,\,\alpha +\beta >3\,, & 
\end{array}%
\right.   \label{MLSE2}
\end{equation}
where $\lambda_3=0_S$ is the zero element of the semigroup satisfying $0_S\lambda_{\alpha}=\lambda_{\alpha}0_S=0_S$. Let $S_{E}=S_0\cup S_1$ be a subset decomposition with
\begin{align}
    S_0&=\{\lambda_0,\lambda_2,\lambda_3\}\,, & S_1&=\{\lambda_1,\lambda_3\}\,.\label{SD2}
\end{align}
and thus satisfying the resonance condition \eqref{SeD}.  After considering a resonant $S_{E}^{\left(2\right)}$-expansion and applying the $0_S$-reduction condition $0_S T_A$=0, we get an expanded hyperalgebra spanned by the set of generators
\begin{eqnarray}
\{J_{a},P_{a},J_{abc},P_{abc},Q_{\alpha a}\}\,.
\end{eqnarray}
The expanded generators are related to the $\mathfrak{osp}\left(1|4\right)$ ones through the semigroup elements as
\begin{equation}
    \begin{tabular}{lll}
%\cline{2-2}\cline{3-3}
\multicolumn{1}{l|}{$\lambda_3$} & \multicolumn{1}{|l}{\cellcolor[gray]{0.8}} & \multicolumn{1}{|l|}{\cellcolor[gray]{0.8}} \\ \hline
\multicolumn{1}{l|}{$\lambda_2$} & \multicolumn{1}{|l}{$P_a,\ P_{abc} $} & \multicolumn{1}{|l|}{\cellcolor[gray]{0.8}} \\ \hline
\multicolumn{1}{l|}{$\lambda_1$} & \multicolumn{1}{|l}{\cellcolor[gray]{0.8}} & \multicolumn{1}{|l|}{$Q_{\alpha a}$} \\ \hline
\multicolumn{1}{l|}{$\lambda_0$} & \multicolumn{1}{|l}{$ J_{a},\ J_{abc}$} & \multicolumn{1}{|l|}{\cellcolor[gray]{0.8}} \\ \hline
\multicolumn{1}{l|}{} & \multicolumn{1}{|l}{$T_{a},\ T_{abc}$} & \multicolumn{1}{|l|}{$\mathcal{G}_{\alpha a}$}
\end{tabular}\label{EXP1}%
\end{equation}
and satisfy the following non-vanishing (anti-)commutation relations:
\begin{eqnarray}
\left[ J_{a},J_{b}\right]  &=&\epsilon _{~ab}^{m}J_{m}\,, \qquad \qquad \ \ \ \, \ \left[
J_{a},P_{b}\right] =\epsilon _{~ab}^{m}P_{m}\,,  \notag \\
\left[ J_{a},J_{bcd}\right]  &=&3\epsilon _{~a\left( b\right. }^{m}J_{\left.
cd\right) m}\,,\qquad \ \ \left[ J_{a},P_{bcd}\right] =3\epsilon _{~a\left(
b\right. }^{m}P_{\left. cd\right) m}\,,  \notag \\
\left[ P_{a},J_{bcd}\right]  &=&3\epsilon _{~a\left( b\right. }^{m}P_{\left.
cd\right) m}\,,\notag\\
\left[ J_{a},Q_{\alpha b}\right] &=&\frac{1}{2}\left( \Gamma _{a}\right) _{%
\text{ }\alpha }^{\beta }Q_{\beta b}+\epsilon _{abc}Q_{\beta }^{~c}\,,
 \notag\\
 \left[ J_{abc},J_{mnk}\right]  &=&-6\left( \eta _{\left( ab\right. }\epsilon
_{~\left. c\right) \left( m\right. }^{l}\eta _{\left. nk\right) }+5\epsilon
_{~\left( m\right. |\left( a\right. }^{l}\delta _{~b}^{d}\eta _{\left.
c\right) |n}\eta _{\left. k\right) d}\right) J_{l}  \notag \\
&&+2\left( 5\epsilon _{~\left( m\right. |\left( a\right. }^{l}\delta
_{~b}^{d}J_{\left. c\right) l|n}\eta _{\left. k\right) d}-\epsilon _{~\left(
m\right. |\left( a\right. }^{l}\eta _{\left. bc\right) }J_{|\left. nk\right)
l}-\epsilon _{~\left( m\right. \left( a\right. |}^{l}J_{\left. bc\right)
l}\eta _{|\left. nk\right) }\right) \,,  \notag \\
\left[ J_{abc},P_{mnk}\right]  &=&-6\left( \eta _{\left( ab\right. }\epsilon
_{~\left. c\right) \left( m\right. }^{l}\eta _{\left. nk\right) }+5\epsilon
_{~\left( m\right. |\left( a\right. }^{l}\delta _{~b}^{d}\eta _{\left.
c\right) |n}\eta _{\left. k\right) d}\right) P_{l}  \notag \\
&&+2\left( 5\epsilon _{~\left( m\right. |\left( a\right. }^{l}\delta
_{~b}^{d}P_{\left. c\right) l|n}\eta _{\left. k\right) d}-\epsilon _{~\left(
m\right. |\left( a\right. }^{l}\eta _{\left. bc\right) }P_{|\left. nk\right)
l}-\epsilon _{~\left( m\right. \left( a\right. |}^{l}P_{\left. bc\right)
l}\eta _{|\left. nk\right) }\right) \,,  \notag \\
\left[ J_{abc},Q_{\alpha d}\right]  &=&\left( \delta _{~d}^{k}\eta _{\left(
ab\right. |}-5\eta _{d\left( a\right. }\delta _{~b|}^{k}\right) \left(
\Gamma _{|\left. c\right) }\right) _{~\alpha }^{\beta }Q_{\beta k}+\eta
_{\left( ab\right. |}\left( \Gamma _{d}\right) _{~\alpha }^{\beta }Q_{\beta
|\left. c\right) }\,,  \notag \\
\left\{ Q_{\alpha a},Q_{\beta b}\right\}  &=&\left( P_{abc}-\frac{4}{3}\eta
_{ab}P_{c}\right) \left( C\Gamma ^{c}\right) _{\alpha \beta }+\frac{5}{3}%
\epsilon _{abc}C_{\alpha \beta }P^{c}+\frac{2}{3}P_{\left( a\right. |}\left(
C\Gamma _{|\left. b\right) }\right) _{\alpha \beta }\,,
 \label{HP}
\end{eqnarray}
where we have used the multiplication law \eqref{MLSE2} and the (anti-)commutation relations of the $\mathfrak{osp}\left(1|4\right)$ superalgebra \eqref{osp14}. The expanded algebra corresponds to the hyper-Poincaré algebra with spin-4 generators,  $\mathfrak{hp}_{(4)}$ \cite{Fuentealba:2015wza}.  One of the advantages of the $S$-expansion procedure is the derivation of the invariant tensor of the expanded (super)algebra in terms of the original one \cite{Izaurieta:2006zz}. Following Theorem VII of ref. \cite{Izaurieta:2006zz}, one can show that the hyper-Poincaré algebra \eqref{HP} admits the following non-vanishing components of an invariant tensor: 
\begin{eqnarray}
\left\langle J_{a}J_{b}\right\rangle &=&\alpha_0\eta _{ab}\,,  \notag \\
\left\langle J_{a}P_{b}\right\rangle &=&\alpha_1\eta _{ab}\,,  \notag \\
\left\langle J_{abc}J_{mnk}\right\rangle &=&2\alpha_0\left(5\eta _{m\left( a\right.
}\eta _{b|n}\eta _{|\left. c\right) k}-3\eta _{\left( ab\right. }\eta
_{|\left. c\right) \left( m\right. }\eta _{\left. nk\right) }\right) \,, \notag \\
\left\langle J_{abc}P_{mnk}\right\rangle &=&2\alpha_1\left(5\eta _{m\left( a\right.
}\eta _{b|n}\eta _{|\left. c\right) k}-3\eta _{\left( ab\right. }\eta
_{|\left. c\right) \left( m\right. }\eta _{\left. nk\right) }\right) \,, \notag \\
\left\langle Q_{\alpha a}Q_{\beta b}\right\rangle &=&%
\alpha_1\left(\frac{4}{3}C_{\alpha \beta }\eta _{ab}-\frac{2}{3}\epsilon _{abc}\left(
C\Gamma ^{c}\right) _{\alpha \beta }\right)\,,  \label{IT2}
\end{eqnarray}
where we have considered
\begin{align}
    \alpha_0&=\lambda_0\mu\,, & \alpha_1&=\lambda_2 \mu\,. \label{redef}
\end{align}
Let $A$ be the gauge connection one-form taking value on $\mathfrak{hp}_{(4)}$,
\begin{eqnarray}
A&=&e^{a}P_{a}+\omega^{a}J_{a}+e^{abc}P_{abc}+\omega^{abc}J_{abc}+\Bar{\psi}^{a}Q_{a}\,,\label{1FHP}
\end{eqnarray}
whose components are the dreibein, the spin connection, the generalized spin-4 vielbein and spin connection and the Majorana spin-$\frac{5}{2}$ gauge fields.  The corresponding curvature two-form reads
\begin{equation}
    F_{\mathfrak{hp}_{(4)}}=\tilde{\mathcal{T}}^{a}P_{a}+\mathcal{R}^{a}J_{a}+\tilde{\mathcal{T}}^{abc}P_{abc}+\mathcal{R}^{abc}J_{abc}+\nabla\Bar{\psi}^{a}Q_{a}\,, \label{2FHP}
\end{equation}
where
\begin{eqnarray}
\tilde{\mathcal{T}}^{a}&=&de^{a}+\epsilon _{~bc}^{a}\omega^{b}e^{c}+30\epsilon^{a}_{\ bc}\omega^{bmn}e^{c}_{\ mn}-\frac{3}{2}i\Bar{\psi}_{b}\Gamma^{a}\psi^{b}\,, \notag \\ 
\mathcal{R}^{a}&=&d\omega^{a}+\frac{1}{2}\epsilon _{~bc}^{a}\omega^{b}\omega^{c}+15\epsilon^{a}_{\ bc}\omega^{bmn}\omega^{c}_{\ mn}\,, \notag \\
\tilde{\mathcal{T}}^{abc}&=&de^{abc}-10\epsilon^{(a}_{\ mn}\omega^{m k|b}e^{c)n}_{\ \ \ k}+3\epsilon^{(a}_{\ mn}e^{m}\omega^{n|bc)}+3\epsilon^{(a}_{\ mn}\omega^{m}e^{n|bc)}+\frac{i}{2}\Bar{\psi}^{(a}\Gamma^{|b}\psi^{c)}\,, \notag \\
\mathcal{R}^{abc}&=&d\omega^{abc}-5\epsilon^{(a}_{\ mn}\omega^{m k|b}\omega^{c)n}_{\ \ \ k}+3\epsilon^{(a}_{\ mn}\omega^{m}\omega^{n|bc)}\,, \notag \\
\nabla\psi^{a}&=&d\psi^{a}+\frac{3}{2}\omega^{b}\Gamma_{b}\psi^{a}-\omega_{b}\Gamma^{a}\psi^{b}-5\omega^{bca}\Gamma_{b}\psi_{c}\,. \label{2FHPb}
\end{eqnarray}
The CS hypergravity action gauge invariant under the $\mathfrak{hp}_{(4)}$ algebra is obtained considering the gauge connection one-form \eqref{1FHP} and the bilinear invariant trace \eqref{IT2} in the general expression of the CS action \eqref{CS}. Omitting the boundary terms, the $\mathfrak{hp}_{(4)}$ CS action reads
   \begin{eqnarray}
     I_{\mathfrak{hp}_{(4)}}&=&\frac{k}{4\pi}\int  \alpha_{0} \mathcal{L}_0+\alpha_{1}\mathcal{L}_{1} \,,  \label{CShp4}
 \end{eqnarray}
where
\begin{eqnarray}
\mathcal{L}_0 &=& \mathcal{L}(\omega) +10\left(d\omega^{abc}-\frac{10}{3}\epsilon^{a}_{\ mn}\omega^{mkb}\omega^{cn}_{\ \ k}+3\epsilon^{a}_{\ mn}\omega^{m}\omega^{nbc}\right)\omega_{abc}\,,\notag \\
\mathcal{L}_{1} &=& 2\mathcal{R}^{a}e_{a}+20\mathcal{R}^{abc}e_{abc}+i\Bar{\psi}_{a}\nabla\psi^{a}\,,
 \label{L0L1}
\end{eqnarray}
with $\mathcal{L}\left(\omega\right)$ being the Lorentz-CS form \eqref{LorCS}. The CS action \eqref{CShp4} is split into two independent sectors proportional to the two arbitrary constants. The term along $\alpha_{0}$ contains the Lorentz CS term \cite{Achucarro:1987vz,Witten:1988hc} together with its spin-$4$ version. Such term can be seen as the exotic counterpart of the $\mathfrak{hp}_{(4)}$ CS gravity action. The term along the $\alpha_1$ defines the hyper-Poincaré gravity action and contains the Einstein-Hilbert term, contributions of the spin-4 fields and a fermionic term. Naturally, one can notice that the CS action reduces to the bosonic Poincaré CS gravity when the spin-$4$ and spin-$\frac{5}{2}$ fields are switch-off. The field equations are given by the vanishing of the curvature 2-form in \eqref{2FHP}, whose components transform covariantly with respect to the hypersymmetry
transformation laws
\begin{eqnarray}
\delta e^{a}&=&3i\Bar{\epsilon}_{b}\Gamma^{a}\psi^{b}\,, \qquad  \ \ \ \delta \omega^{a}=0\,,  \notag \\
\delta e^{abc}&=&-i\Bar{\epsilon}^{a}\Gamma^{b}\psi^{c}\,, \qquad  \delta \omega^{abc}=0\,, \notag \\
\delta \psi^{a}&=&d\epsilon^{a}+\frac{3}{2}\omega^{b}\Gamma_{b}\epsilon^{a}-\omega_{b}\Gamma^{a}\epsilon^{b}-5\omega^{bca}\Gamma_{b}\epsilon_{c}\,.
\end{eqnarray}
where $\epsilon^{a}$ is the fermionic gauge parameters.

In absence of spin-4 gauge fields, we recover the hypergravity theory proposed in \cite{Fuentealba:2015jma} whose action coincides with the one presented by Aragone and Deser in \cite{Aragone:1983sz}. However our action, with vanishing spin-4 gauge fields, differs from the one presented in \cite{Aragone:1983sz} in the hypersymmetry transformation rule for the spin connection. By construction, as in supergravity \cite{Banados:1996hi}, the gauge symmetry
\begin{eqnarray}
\delta e^{a}&=&D_{\omega}\rho^{a}-\epsilon^{abc}\sigma_{b}e_{c}+3i\Bar{\epsilon}_{b}\Gamma^{a}\psi^{b}\,, \notag  \\
\delta \omega^{a}&=&D_{\omega}\sigma^{a}\,,  \notag \\
\delta \psi^{a}&=&d\epsilon^{a}+\frac{3}{2}\omega^{b}\Gamma_{b}\epsilon^{a}-\omega_{b}\Gamma^{a}\epsilon^{b}-5\omega^{bca}\Gamma_{b}\epsilon_{c}\,.
\end{eqnarray}
closes off-shell, without the need of auxiliary fields. Here, $D_{\omega}=d+\omega$ denotes the covariant derivative, while $\rho^{a}$ and $\sigma^{a}$ denote the bosonic gauge parameters related to the $J_{a}$ and $P_{a}$ generators, respectively.
%%%%%%%%%%%%%%%%%%%%%%%%%%%%%%%%%%%%%%%%%%%%%%%%%%%%%%%%%%%%%%%%%%%%%%%%%%%%%%%%%%%%%%%%%%%%%%%%%%%
\subsection{Hyper-Maxwell gravity}
A hyper-Maxwell algebra with spin-4 generators, denoted as $\mathfrak{hm}_{(4)}$ in \cite{Caroca:2021bjo}, can alternatively be obtained by expanding the $\mathfrak{osp}\left(1|4\right)$ superalgebra considering $S_{E}^{\left(4\right)}=\{\lambda_0,\lambda_1,\lambda_2,\lambda_3,\lambda_4,\lambda_5\}$ as the relevant semigroup. The elements of the $S_{E}^{\left(4\right)}$ semigroup satisfy the following multiplication law:
\begin{equation}
\lambda _{\alpha }\lambda _{\beta }=\left\{ 
\begin{array}{lcl}
\lambda _{\alpha +\beta }\,\,\,\, & \mathrm{if}\,\,\,\,\alpha +\beta \leq
5\,, &  \\ 
\lambda _{5}\,\,\, & \mathrm{if}\,\,\,\,\alpha +\beta >5\,, & 
\end{array}%
\right.   \label{MLSE4}
\end{equation}
where $\lambda_5$ plays the role of the zero element of the semigroup. Before applying the expansion, we require a resonant subset decomposition $S_{E}^{\left(4\right)}=S_0\cup S_1$ with
\begin{align}
    S_0&=\{\lambda_0,\lambda_2,\lambda_4,\lambda_5\}\,, & S_1&=\{\lambda_1,\lambda_3,\lambda_5\}\,.\label{SD3}
\end{align}
Such decomposition is said to be resonant since it satisfies the same algebraic structure than the subspaces of the $\mathfrak{osp}\left(1|4\right)$ superalgebra. An expanded hyperalgebra is obtained after considering a resonant $S_{E}^{\left(4\right)}$-expansion and performing a $0_S$-reduction. The expanded algebra is spanned by the set of generators
\begin{equation}
  \lbrace J_a,P_a,Z_a,J_{abc},P_{abc},Z_{abc},Q_{\alpha a},\Sigma_{\alpha a} \rbrace \,,  
\end{equation}
which are related to the $\mathfrak{osp}\left(1|4\right)$ ones through the semigroup elements as
\begin{equation}
    \begin{tabular}{lll}
%\cline{2-2}\cline{3-3}
\multicolumn{1}{l|}{$\lambda_5$} & \multicolumn{1}{|l}{\cellcolor[gray]{0.8}} & \multicolumn{1}{|l|}{\cellcolor[gray]{0.8}} \\ \hline
\multicolumn{1}{l|}{$\lambda_4$} & \multicolumn{1}{|l}{$Z_a,\ Z_{abc} $} & \multicolumn{1}{|l|}{\cellcolor[gray]{0.8}} \\ \hline
\multicolumn{1}{l|}{$\lambda_3$} & \multicolumn{1}{|l}{\cellcolor[gray]{0.8}} & \multicolumn{1}{|l|}{$\Sigma_{\alpha a}$} \\ \hline
\multicolumn{1}{l|}{$\lambda_2$} & \multicolumn{1}{|l}{$P_a,\ P_{abc} $} & \multicolumn{1}{|l|}{\cellcolor[gray]{0.8}} \\ \hline
\multicolumn{1}{l|}{$\lambda_1$} & \multicolumn{1}{|l}{\cellcolor[gray]{0.8}} & \multicolumn{1}{|l|}{$Q_{\alpha a}$} \\ \hline
\multicolumn{1}{l|}{$\lambda_0$} & \multicolumn{1}{|l}{$ J_{a},\ J_{abc}$} & \multicolumn{1}{|l|}{\cellcolor[gray]{0.8}} \\ \hline
\multicolumn{1}{l|}{} & \multicolumn{1}{|l}{$T_{a},\ T_{abc}$} & \multicolumn{1}{|l|}{$\mathcal{G}_{\alpha a}$}
\end{tabular}\label{EXP1b}%
\end{equation}
The explicit (anti-)commutation relations for the expanded hyperalgebra are obtained by considering the multiplication law \eqref{MLSE4} of the semigroup and the original (anti-)commutation relations of the $\mathfrak{osp}\left(1|4\right)$ superalgebra. The expanded generators satisfy the (anti-)commutation relations of the hyper-Poincaré algebra \eqref{HP} along the following ones:
\begin{eqnarray}
\left[ J_{a},Z_{b}\right]  &=&\epsilon _{~ab}^{m}Z_{m}\,, \qquad \qquad \qquad \ \ \, \left[
P_{a},P_{b}\right] =\epsilon _{~ab}^{m}Z_{m}\,,  \notag \\
\left[ J_{a},Z_{bcd}\right] &=&3\epsilon _{~a\left(
b\right. }^{m}Z_{\left. cd\right) m}\,,  \qquad \qquad \left[ Z_{a},J_{bcd}\right]  =3\epsilon _{~a\left( b\right. }^{m}Z_{\left.
cd\right) m}\,, \notag \\
\left[ P_{a},P_{bcd}\right] &=&3\epsilon _{~a\left(
b\right. }^{m}Z_{\left. cd\right) m}\,, \notag \\
\left[ J_{a},\Sigma _{\alpha b}\right]  &=&\frac{1}{2}\left( \Gamma
_{a}\right) _{\text{ }\alpha }^{\beta }\Sigma _{\beta b}+\epsilon
_{abc}\Sigma _{\beta }^{~c}\,,  \notag \\
\left[ P_{a},Q_{\alpha b}\right]  &=&\frac{1}{2}\left( \Gamma _{a}\right) _{%
\text{ }\alpha }^{\beta }\Sigma _{\beta b}+\epsilon _{abc}\Sigma _{\beta
}^{~c}\,, \notag \\
\left[ J_{abc},Z_{mnk}\right]  &=&-6\left( \eta _{\left( ab\right. }\epsilon
_{~\left. c\right) \left( m\right. }^{l}\eta _{\left. nk\right) }+5\epsilon
_{~\left( m\right. |\left( a\right. }^{l}\delta _{~b}^{d}\eta _{\left.
c\right) |n}\eta _{\left. k\right) d}\right) Z_{l} \notag \\
&&+2\left( 5\epsilon _{~\left( m\right. |\left( a\right. }^{l}\delta
_{~b}^{d}Z_{\left. c\right) l|n}\eta _{\left. k\right) d}-\epsilon _{~\left(
m\right. |\left( a\right. }^{l}\eta _{\left. bc\right) }Z_{|\left. nk\right)
l}-\epsilon _{~\left( m\right. \left( a\right. |}^{l}Z_{\left. bc\right)
l}\eta _{|\left. nk\right) }\right) \,,  \notag \\
\left[ P_{abc},P_{mnk}\right]  &=&-6\left( \eta _{\left( ab\right. }\epsilon
_{~\left. c\right) \left( m\right. }^{l}\eta _{\left. nk\right) }+5\epsilon
_{~\left( m\right. |\left( a\right. }^{l}\delta _{~b}^{d}\eta _{\left.
c\right) |n}\eta _{\left. k\right) d}\right) Z_{l}  \notag \\
&&+2\left( 5\epsilon _{~\left( m\right. |\left( a\right. }^{l}\delta
_{~b}^{d}Z_{\left. c\right) l|n}\eta _{\left. k\right) d}-\epsilon _{~\left(
m\right. |\left( a\right. }^{l}\eta _{\left. bc\right) }Z_{|\left. nk\right)
l}-\epsilon _{~\left( m\right. \left( a\right. |}^{l}Z_{\left. bc\right)
l}\eta _{|\left. nk\right) }\right) \,, \notag \\
\left[ J_{abc},\Sigma _{\alpha d}\right]  &=&\left( \delta _{~d}^{k}\eta
_{\left( ab\right. |}-5\eta _{d\left( a\right. }\delta _{~b|}^{k}\right)
\left( \Gamma _{|\left. c\right) }\right) _{~\alpha }^{\beta }\Sigma _{\beta
k}+\eta _{\left( ab\right. |}\left( \Gamma _{d}\right) _{~\alpha }^{\beta
}\Sigma _{\beta |\left. c\right) }\,,  \notag \\
\left[ P_{abc},Q_{\alpha d}\right]  &=&\left( \delta _{~d}^{k}\eta _{\left(
ab\right. |}-5\eta _{d\left( a\right. }\delta _{~b|}^{k}\right) \left(
\Gamma _{|\left. c\right) }\right) _{~\alpha }^{\beta }\Sigma _{\beta
k}+\eta _{\left( ab\right. |}\left( \Gamma _{d}\right) _{~\alpha }^{\beta
}\Sigma _{\beta |\left. c\right) }\,, \notag \\
\left\{ Q_{\alpha a},\Sigma _{\beta b}\right\}  &=&\left( Z_{abc}-\frac{4}{3}%
\eta _{ab}Z_{c}\right) \left( C\Gamma ^{c}\right) _{\alpha \beta }+\frac{5}{3%
}\epsilon _{abc}C_{\alpha \beta }Z^{c}+\frac{2}{3}Z_{\left( a\right.
|}\left( C\Gamma _{|\left. b\right) }\right) _{\alpha \beta }\,. \label{HM}
\end{eqnarray}%
This expanded algebra corresponds to one of the three inequivalent hypersymmetric extensions of the Maxwell algebra presented in \cite{Caroca:2021bjo}, which allowed to couple consistently Maxwell CS gravity theory with massless spin-$\frac{5}{2}$ gauge field. The algebra \eqref{HM}, denoted as $\mathfrak{hm}_{(4)}$, was first obtained as an Inönü-Wigner contraction of the $\mathfrak{osp}\left(1|4\right)^{2}\otimes\mathfrak{sp}\left(4\right)$ superalgebra \cite{Caroca:2021bjo}. Let us note that the present hyper-Maxwell algebra is characterized by the presence of a second fermionic generator $\Sigma_{\alpha a}$ analogously to the minimal Maxwell superalgebra \cite{Bonanos:2009wy,deAzcarraga:2014jpa,Concha:2014tca,Concha:2019mxx}. The presence of two fermionic charges, unlike the two other hyper-Maxwell algebras introduced in \cite{Caroca:2021bjo}, allows to express the translational generators $P_a$ as bilinear expressions of fermionic generators $Q_{\alpha a}$ ensuring the construction of a consistent hypergravity CS action. 

The $\mathfrak{hm}_{(4)}$ CS action is constructed from the gauge field 
\begin{equation}
    A=e^{a}P_{a}+\omega^{a}J_{a}+k^{a}Z_{a}+e^{abc}P_{abc}+\omega^{abc}J_{abc}+k^{abc}Z_{abc}+\Bar{\psi}^{a}Q_{a}+\Bar{\xi}^{a}\Sigma_{a}\,,\label{1FHM}
\end{equation}
where $k_{a}$ denotes the gravitational Maxwell field, $k^{abc}$ is an additional spin-4 generators and $\xi^{a}$ is a second Majorana spin-$\frac{5}{2}$ gauge field.  The corresponding curvature two-form reads \cite{Caroca:2021bjo}
\begin{equation}
    F_{\mathfrak{hm}_{(4)}}=\tilde{\mathcal{T}}^{a}P_{a}+\mathcal{R}^{a}J_{a}+\tilde{\mathcal{F}}^{a}Z_{a}+\tilde{\mathcal{T}}^{abc}P_{abc}+\mathcal{R}^{abc}J_{abc}+\tilde{\mathcal{F}}^{abc}Z_{abc}+\nabla\Bar{\psi}^{a}Q_{a}+\nabla\Bar{\xi}^{a}\Sigma_{a}\,, \label{2FHM}
\end{equation}
where $\tilde{\mathcal{T}}^{a}$, $\mathcal{R}^{a}$, $\tilde{\mathcal{T}}^{abc}$, $\mathcal{R}^{abc}$ and $\nabla\psi^{a}$ coincide with the hyper-Poincaré ones \eqref{2FHPb} and
\begin{eqnarray}
\tilde{\mathcal{F}}^{a}&=&dk^{a}+\epsilon _{~bc}^{a}\omega^{b}k^{c}+\frac{1}{2}
\epsilon _{~bc}^{a}e^{b}e^{c}+30\epsilon^{a}_{\ bc}\omega^{bmn}k^{c}_{\ mn}+15\epsilon^{a}_{\ bc}e^{bmn}e^{c}_{\ mn}-3i\Bar{\psi}_{b}\Gamma^{a}\xi^{b}\,, \notag \\
\tilde{\mathcal{F}}^{abc}&=&dk^{abc}-10\epsilon^{(a}_{\ mn}\omega^{m k|b}k^{c)n}_{\ \ \ k}-5\epsilon^{(a}_{\ mn}e^{m k|b}e^{c)n}_{\ \ \ k}+3\epsilon^{(a}_{\ mn}\omega^{m}k^{n|bc)}+3\epsilon^{(a}_{\ mn}k^{m}\omega^{n|bc)} \notag \\
&&+3\epsilon^{(a}_{\ mn}e^{m}e^{n|bc)}+i\Bar{\psi}^{(a}\Gamma^{|b}\xi^{c)}\,, \notag \\
\nabla\xi^{a}&=&d\xi^{a}+\frac{3}{2}\omega^{b}\Gamma_{b}\xi^{a}-\omega_{b}\Gamma^{a}\xi^{b}-5\omega^{bca}\Gamma_{b}\xi_{c}+\frac{3}{2}e^{b}\Gamma_{b}\psi^{a}-e_{b}\Gamma^{a}\psi^{b}-5e^{bca}\Gamma_{b}\psi_{c}\,. \label{2FHMb}
\end{eqnarray}
From Theorem VII of ref. \cite{Izaurieta:2006zz}, one can see that the $\mathfrak{hm}_{(4)}$ hyper-algebra admits the non-vanishing components of the hyper-Poincaré invariant tensor \eqref{IT2} along with the following ones
\begin{eqnarray}
 \langle P_{a}P_{b} \rangle&=&\alpha_{2}\eta_{ab}\,,\notag \\
 \langle J_{a}Z_{b} \rangle&=&\alpha_{2}\eta_{ab}\,, \notag \\
 \langle J_{abc}Z_{mnk} \rangle&=&2\alpha_{2}\left(5 \eta_{m(a}\eta_{b|n}\eta_{|c)k}-3\eta_{(ab|}\eta_{|c)}(m|\eta_{|nk)} \right)\,,\notag \\
 \langle P_{abc}P_{mnk} \rangle&=&2\alpha_{2}\left(5 \eta_{m(a}\eta_{b|n}\eta_{|c)k}-3\eta_{(ab|}\eta_{|c)}(m|\eta_{|nk)} \right)\,,\notag \\
 \langle Q_{\alpha a}\Sigma_{\beta b}\rangle &=&\alpha_{2}\left(\frac{4}{3}C_{\alpha \beta} \eta_{ab}-\frac{2}{3}\epsilon_{abc}\left(C\Gamma^{c}\right)_{\alpha\beta}\right)\,.\label{IT3}
\end{eqnarray}
 Here $\alpha _{0}$, $\alpha _{1}$ and $\alpha _{2}$ are arbitrary constants which are related to the $\mathfrak{osp}\left(1|4\right)$ constants though the $S_{E}^{\left(4\right)}$ semigroup elements as
 \begin{align}
    \alpha_0&=\lambda_0\mu\,, & \alpha_1&=\lambda_2 \mu \,, & \alpha_2&=\lambda_4 \mu\,. \label{redef2}
\end{align}
The $\mathfrak{hm}_{(4)}$ CS hypergravity action is obtained by replacing the gauge connection one-form \eqref{1FHM} and the invariant tensor \eqref{IT2} along \eqref{IT3} in the general expression for a CS action \eqref{CS}. Then, the CS action reduces, up to a surface term, to
   \begin{eqnarray}
     I_{\mathfrak{hm}_{(4)}}&=&\frac{k}{4\pi}\int \alpha_{0} \mathcal{L}_{0} +\alpha_{1} \mathcal{L}_{1}+\alpha_2 \mathcal{L}_{2}\,,  \label{CSHM}
 \end{eqnarray}
where
\begin{eqnarray}
\mathcal{L}_{2}&=&2\mathcal{R}^{a}k_{a}+e^{a}\mathcal{T}_{a}+20\mathcal{R}^{abc}k_{abc}+10 e^{abc}\mathcal{T}_{abc}+2i\Bar{\psi}_{a}\nabla\xi^{a}+2i\Bar{\xi}_{a}\nabla\psi^{a}\,, \label{L2}
\end{eqnarray}
while $\mathcal{L}_{0}$ and $\mathcal{L}_{1}$ are the respective exotic hypergravity and the hyper-Poincaré Lagrangian defined in \eqref{L0L1}. Here $\mathcal{T}^a$ and $\mathcal{T}^{abc}$ are given by
\begin{eqnarray}
\mathcal{T}^{a}&=&de^{a}+\epsilon _{~bc}^{a}\omega^{b}e^{c}+30\epsilon^{a}_{\ bc}\omega^{bmn}e^{c}_{\ mn}\,,  \nonumber \\ 
\mathcal{T}^{abc}&=&de^{abc}-10\epsilon^{(a}_{\ mn}\omega^{m k|b}e^{c)n}_{\ \ \ k}+3\epsilon^{(a}_{\ mn}e^{m}\omega^{n|bc)}+3\epsilon^{(a}_{\ mn}\omega^{m}e^{n|bc)}\,.
\label{Ts}
\end{eqnarray}
It is interesting to note that the most general hypergravity CS action for the $\mathfrak{hm}_{(4)}$ algebra contains both $\mathcal{L}_0$ and $\mathcal{L}_{1}$ Lagrangian of the hyper-Poincaré gravity theory. The new gauge fields of the hyper-Maxwell algebra contribute exclusively along the $\alpha_2$ constant which requires the presence of a second spinor gauge field $\xi^{a}$ in order to ensure the proper coupling of Maxwell CS gravity with massless spin-$\frac{5}{2}$ gauge fields \cite{Caroca:2021bjo}. One can note that, for $\alpha_2\neq0$, the fields equations are given by the vanishing of the curvature two-forms \eqref{2FHM}.
%%%%%%%%%%%%%%%%%%%%%%%%%%%%%%%%%%%%%%%%%%%%%%%%%%%%%%%%%%%%%%%%%%%%%%%%%%%%%%%%%%%%%%%%%%%%%%%%%%%
\subsection{Generalized hyper-Poincaré gravity}
A generalized hyper-Poincaré algebra can be obtained by considering the $S$-expansion of the $\mathfrak{osp}\left(1|4\right)$ superalgebra with $S_{E}^{\left(2N\right)}=\{\lambda_0,\lambda_1,\cdots,\lambda_{2N},\lambda_{2N+1}\}$ as the relevant semigroup whose elements satisfy
\begin{equation}
\lambda _{\alpha}\lambda _{\beta} =\left\{
\begin{array}{lcl}
\lambda _{\alpha+\beta}\,\,\,\, & \mathrm{if}\,\,\,\,\alpha+\beta \leq 2N\,, &
\\
\lambda _{2N+1}\,\, & \mathrm{if}\,\,\,\,\alpha+\beta > 2N\,, &
\end{array}
\right.  \label{MLSEN}
\end{equation}
where $\lambda_{2N+1}=0_S$ is the zero element of the semigroup. Let $S_{E}^{\left(2N\right)}=S_0\cup S_1$ be a semigroup decomposition with
\begin{eqnarray}
S_0&=&\lbrace \lambda_{2i}, \ i=0,\ldots,N\rbrace \cup\lambda_{2N+1}\,,\notag\\
S_1&=&\lbrace \lambda_{2i+1}, \ i=0,\ldots,N-1\rbrace\cup\lambda_{2N+1}\,,
\end{eqnarray}
which is said to be resonant since it satisfies the same structure than the subspace decomposition of the original $\mathfrak{osp}\left(1|4\right)$ superalgebra \eqref{SD}. Then, after performing a resonant $S_{E}^{\left(2N\right)}$-expansion to the $\mathfrak{osp}\left(1|4\right)$ superalgebra \eqref{osp14} and considering a $0_S$-reduction, we find a new hyperalgebra. The expanded generators are related to the original ones through the semigroup elements as follows:
\begin{align}
    J^{(i)}_{a}&=\lambda_{2i} T_{a}\,, & J^{(i)}_{abc}&=\lambda_{2i} T_{abc}\,, & Q^{(i)}_{\alpha a}&=\lambda_{2i+1}\mathcal{G}_{\alpha a}\,. \label{EG2}
\end{align}
The expanded (anti-)commutators are obtained considering the multiplication law of the semigroup $S_{E}^{\left(2N\right)}$ \eqref{MLSEN} along the original (anti-)commutation relations \eqref{osp14}. Indeed, we find
\begin{align}
\left[ J^{(i)}_{a},J^{(j)}_{b}\right] &=\epsilon _{~ab}^{m}T^{(i+j)}_{m}\,,  \notag \\
\left[ J^{(i)}_{a},J^{(j)}_{bcd}\right] &=3\epsilon _{~a\left( b\right. }^{m}J^{(i+j)}_{\left.
cd\right) m}\,,  \notag \\
\left[ J^{(i)}_{a},Q^{(j)}_{\alpha b}\right] &=\frac{1}{2}\left( \Gamma
_{a}\right) _{\ \alpha }^{\beta }Q^{(i+j)}_{\beta b}+\epsilon _{abc}%
Q^{~c\,(i+j)}_{\beta }\,,  \notag \\
\left[ J^{(i)}_{abc},J^{(j)}_{mnk}\right] &=-6\left( \eta _{\left( ab\right. }\epsilon
_{~\left. c\right) \left( m\right. }^{l}\eta _{\left. nk\right) }+5\epsilon
_{~\left( m\right. |\left( a\right. }^{l}\delta _{~b}^{d}\eta _{\left.
c\right) |n}\eta _{\left. k\right) d}\right) J^{(i+j)}_{l} \notag  \\
&+2\left( 5\epsilon _{~\left( m\right. |\left( a\right. }^{l}\delta
_{~b}^{d}J^{(i+j)}_{\left. c\right) l|n}\eta _{\left. k\right) d}-\epsilon _{~\left(
m\right. |\left( a\right. }^{l}\eta _{\left. bc\right) }J^{(i+j)}_{|\left. nk\right)
l}-\epsilon _{~\left( m\right. \left( a\right. |}^{l}J^{(i+j)}_{\left. bc\right)
l}\eta _{|\left. nk\right) }\right) \,,  \notag \\
\left[ J^{(i)}_{abc},Q^{(j)}_{\alpha d}\right] &=\left( \delta _{~d}^{k}\eta
_{\left( ab\right. |}-5\eta _{d\left( a\right. }\delta _{~b|}^{k}\right)
\left( \Gamma _{|\left. c\right) }\right) _{~\alpha }^{\beta }Q^{(i+j)}%
_{\beta k}+\eta _{\left( ab\right. |}\left( \Gamma _{d}\right) _{~\alpha
}^{\beta }Q^{(i+j)}_{\beta |\left. c\right) }\,,  \notag \\
\left\{ Q^{(i)}_{\alpha a},Q^{(j)}_{\beta b}\right\} &=\left(
J^{(i+j+1)}_{abc}-\frac{4}{3}\eta _{ab}J^{(i+j+1)}_{c}\right) \left( C\Gamma ^{c}\right)
_{\alpha \beta }+\frac{5}{3}\epsilon _{abc}C_{\alpha \beta }J^{c\,(i+j+1)}+\frac{2}{3}%
J^{(i+j+1)}_{\left( a\right. |}\left( C\Gamma _{|\left. b\right) }\right) _{\alpha
\beta }\,.  \label{GHP}
\end{align}%
The expanded hyperalgebra, which we shall denote as $\mathfrak{ghp}^{(N)}$, generalizes the hyper-Poincaré algebra \cite{Fuentealba:2015wza} and corresponds to a hypersymmetric extension of the generalized Poincaré algebra \cite{Edelstein:2006se,Izaurieta:2009hz,Concha:2014zsa}. One can note that both spin-2 and spin-4 generators ($J_{a}^{\left(i\right)}$ and $J_{abc}^{\left(i\right)}$) act non-trivially on the fermionic charges $Q_{\alpha a}^{\left(j\right)}$ if $i+j<N$. On the contrary, the bosonic generators act as central charges. The hyper-Poincaré algebra \eqref{HP} is recovered for $N=1$, while the $\mathfrak{ghp}^{(2)}$ reproduces the hyper-Maxwell one \eqref{HM}. For $N=3$, the hyper-algebra \eqref{GHP} corresponds to a hypersymmetric extension of the generalized Maxwell algebra \cite{Concha:2020sjt}, also denoted as $\mathfrak{B}_{5}$ \cite{Edelstein:2006se,Izaurieta:2009hz}. Interestingly, the generalization can be extended for infinite $N$ leading to an infinite-dimensional generalization of the hyper-Poincaré algebra. In such case, the semigroup $S_{E}^{(\infty)}$ does not contain zero element and the infinite-dimensional hyper-algebra $\mathfrak{ghp}^{(\infty)}$ does not admit abelian commutators neither central charges in \eqref{GHP}.

The non-vanishing components of the invariant tensor for the $\mathfrak{ghp}^{\left(N\right)}$ hyperalgebra can be obtained from the $\mathfrak{osp}\left(1|4\right)$ ones considering the semigroup $S_{E}^{(N)}$ and following the Theorem VII of ref. \cite{Izaurieta:2006zz}. Indeed, one can see that the generalized hyper-poincaré algebra admits the following invariant bilinear trace:
\begin{eqnarray}
\left\langle J^{(i)}_{a}J^{(j)}_{b}\right\rangle &=&\alpha_{i+j}\eta _{ab}\,,  \notag \\
\left\langle J^{(i)}_{abc}J^{(j)}_{mnk}\right\rangle &=&2\alpha_{i+j}\left(5\eta _{m\left( a\right.
}\eta _{b|n}\eta _{|\left. c\right) k}-3\eta _{\left( ab\right. }\eta
_{|\left. c\right) \left( m\right. }\eta _{\left. nk\right) }\right) \,, \notag \\
\left\langle Q^{(i)}_{\alpha a}Q^{(j)}_{\beta b}\right\rangle &=&%
\alpha_{i+j}\left(\frac{4}{3}C_{\alpha \beta }\eta _{ab}-\frac{2}{3}\epsilon _{abc}\left(
C\Gamma ^{c}\right) _{\alpha \beta }\right)\,,  \label{ITGHP}
\end{eqnarray}
where $\alpha_{i}$ is an arbitrary constant related to the $\mathfrak{osp}\left(1|4\right)$ one through the semigroup elements as
\begin{eqnarray}
\alpha_{i}&=&\lambda_{2i}\mu\,,
\end{eqnarray}
for $0\leq i\leq N$. In particular, $\alpha_{i+j}$ is zero for $i+j\geq N+1$. An hypergravity CS action for the $\mathfrak{ghp}^{\left(N\right)}$ hyperalgebra can be constructed considering the non-vanishing components of the invariant tensor \eqref{ITGHP} and the gauge connection one-form:
\begin{eqnarray}
A&=&\sum_{i=0}^{N}\left(\omega^{a\,(i)}J^{(i)}_{a}+\omega^{abc\,(i)}J^{(i)}_{abc}\right)+\sum_{i=0}^{N-1}\left( \bar{\psi}^{a\,(i)}Q^{(i)}_{a}\right)\,,\label{1FGHP}
\end{eqnarray}
in the general expression of the CS action \eqref{CS}, that yields
\begin{eqnarray}
I_{\mathfrak{ghp}^{\left(N\right)}}&=&\frac{k}{4\pi}\int \sum_{i=0}^{N}\alpha_{i}\mathcal{L}_{i}=\frac{k}{4\pi}\int \alpha_{0}\mathcal{L}_{0}+\alpha_1\mathcal{L}_{1}+\alpha_2\mathcal{L}_{2}+\cdots+\alpha_{N}\mathcal{L}_{N}\,,
\end{eqnarray}
where
\begin{align}
\mathcal{L}_{i}&= \omega^{(j)}_{a}d\omega^{a\,(k)}\delta_{j+k}^{i}+ \frac{1}{3}\epsilon _{~bc}^{a}\omega^{(j)}_{a}\omega^{b\,(k)}\omega^{c\,(l)}\delta_{j+k+l}^{i}+\bar{\psi}^{(j)}_{a}\nabla\psi^{a\,(k)}\delta_{j+k+1}^{i} \notag\\
&+\left(d\omega^{abc\,(j)}\delta_{j+l}^{i}-\frac{10}{3}\epsilon^{a}_{\ mn}\omega^{mpb\,(j)}\omega^{cn\,(k)}_{\ \ p}\delta_{j+k+l}^{i}+3\epsilon^{a}_{\ mn}\omega^{m\,(j)}\omega^{nbc\,(k)}\delta_{j+k+l}^{i}\right)\omega^{(l)}_{abc} \,. \label{LI}
\end{align}
Here, the covariant derivative of the spinor 1-form for the $\mathfrak{ghp}^{\left(N\right)}$ superalgebra reads
\begin{align}
    \nabla\psi^{a\,(i)}&=d\psi^{a\,(i)}+\frac{3}{2}\left(\omega^{b\,(j)}\Gamma_{b}\psi^{a\,(k)}-\omega^{(j)}_{b}\Gamma^{a}\psi^{b\,(k)}-5\omega^{bca\,(j)}\Gamma_{b}\psi^{(k)}_{c}\right)\delta_{j+k}^{i}\,.
\end{align}
Let us note that $\mathcal{L}_{0}$ is the exotic Lagrangian \cite{Witten:1988hc} coupled to spin-4 gauge field defined in \eqref{L0L1}. On the other hand, $\mathcal{L}_{1}$ and $\mathcal{L}_{2}$ are the respective hyper-Poincaré and hyper-Maxwell Lagrangian obtained in \eqref{L0L1} and \eqref{L2}, respectively. New generalizations of the hyper-Poincaré gravity theory appear for $N\geq3$, corresponding to hypersymmetric extensions of the $\mathfrak{B}_{N+2}$ CS gravity \cite{Concha:2014vka,Concha:2014zsa}. Thus, the hypergravity CS action gauge invariant under the generalized hyper-Poincaré algebra can be written as the sum of diverse hypergravity CS Lagrangians. Interestingly, the $\mathfrak{ghp}^{\left(N\right)}$ hypergravity CS action can alternatively be recovered from the $\mathfrak{osp}\left(1|4\right)$ CS action \eqref{ospCS} by expressing the $\mathfrak{ghp}^{\left(N\right)}$ gauge field in terms of the $\mathfrak{osp}\left(1|4\right)$ ones as
\begin{align}
    \omega^{a\,\left(i\right)}&=\lambda_{2i} W^{a}\,, & \omega^{abc\,\left(i\right)}&=\lambda_{2i}W^{abc}\,, &\psi^{a\,\left(i\right)}&=\lambda_{2i+1}\Psi^{a}\,. \label{gfresc}
\end{align}
One can notice that the field equations for the $\mathfrak{ghp}^{\left(i\right)}$ hypergravity theory are given by the vanishing of the curvature two-forms:
\begin{eqnarray}
F^{a\,\left(i\right)}&=&d\omega^{a\,(i)}+\left(\frac{1}{2}\epsilon^{a}_{\ bc}\omega^{b\,(j)}\omega^{c\,(k)}+15\epsilon^{a}_{bc}\omega^{bmn\,(j)}\omega^{c\,(k)}_{\ mn}\right)\delta_{j+k}^{i}-\frac{3}{2}i\bar{\psi}_{b\,(j)}\Gamma^{a}\psi^{b\,(k)}\delta_{j+k+1}^{i}\,, \notag \\
    F^{abc\,\left(i\right)}&=&d\omega^{abc\,(i)}-\left(5\epsilon^{(a}_{\ mn}\omega^{m p|b\,(j)}\omega^{c)n\,(k)}_{\ \ \ p}-3\epsilon^{(a}_{\ mn}\omega^{m\,(j)}\omega^{n|bc)\,(k)}\right)\delta_{j+k}^{i}\notag\\
    &&+\frac{i}{2}\bar{\psi}^{(a\,(j)}\Gamma^{|b}\psi^{c)\,(k)}\delta_{j+k+1}^{i}\,, \notag \\
    \nabla\psi^{a\,\left(i\right)}&=&d\psi^{a\,(i)}+\left(\frac{3}{2}\omega^{b\,(j)}\Gamma_{b}\psi^{a\,(k)}-\omega_{b}^{(j)}\Gamma^{a}\psi^{b\,(k)}-5\omega^{bca\,(j)}\Gamma_{b}\psi_{c}^{(k)}\right)\delta_{j+k}^{i}\,. \label{2FGHP}
\end{eqnarray}
%%%%%%%%%%%%%%%%%%%%%%%%%%%%%%%%%%%%%%%%%%%%%%%%%%%%%%%%%%%%%%%%%%%%%%%%%%%%%%%%%%%%%%%%%%%%%%%%%%%

\section{Generalized hyper-AdS gravity theory}\label{sec4}
It is interesting to note that one can obtain a different family of algebras considering another semigroup. Indeed, the $S_{\mathcal{M}}^{\left(N\right)}$  semigroup allows to obtain a generalized AdS algebra\footnote{The generalized AdS algebra is also denoted as $\mathfrak{C}_{k}$ in the literature and has been studied in diverse context \cite{Concha:2016kdz,Paixao:2019usi,Cardenas:2022iep}.} \cite{Salgado:2014qqa} from the $\mathfrak{so}\left(2,1\right)$ \cite{Caroca:2017onr}. As we shall see, such expansion behavior will be inherited in higher spin algebras. Here, we present a new generalized hyper-AdS algebra by expanding the $\mathfrak{osp}\left(1|4\right)$ superalgebra with $S_{\mathcal{M}}^{\left(2N\right)}$ as the relevant semigroup. We then extend the procedure to the derivation of the bilinear invariant trace required to construct hypergravity CS actions.
\subsection{Hyper-AdS gravity}
In absence of fermionic generators, the $\mathfrak{so}\left(2,2\right)$ algebra can be obtained from the $\mathfrak{so}\left(2,1\right)$ algebra using the $S_{\mathcal{M}}^{\left(1\right)}$ semigroup (see diagram \eqref{diagram1}). However in presence of supersymmetry, as it was shown in \cite{Caroca:2019dds}, the semigroup required to obtain the AdS superalgebra from the  $\mathfrak{osp}\left(2|1\right)$ superalgebra is the $S_{L}^{\left(1\right)}$ one. Here, following the supersymmetric case, we consider the $S$-expansion of the $\mathfrak{osp}\left(1|4\right)$ superalgebra using the $S_{L}^{\left(1\right)}$ semigroup in order to obtain a hyper-AdS algebra defined in three spacetime dimensions. In particular, we show that the $\mathfrak{sp}\left(4\right)\otimes\mathfrak{osp}\left(1|4\right)$ superalgebra appears by expanding the $\mathfrak{osp}\left(1|4\right)$ superalgebra with $S_{L}^{\left(1\right)}$. We also show that there are two different changes of basis allowing to obtain a hyper-AdS algebra which in the vanishing cosmological constant limit $\ell\rightarrow\infty$ reproduce two inequivalent hyper-Poincaré algebras with and without spin-4 generators \cite{Fuentealba:2015wza}.

Let us consider $S_{L}^{\left( 1\right)
}=\lbrace \lambda _{0},\lambda _{1},\lambda _{2}\rbrace $ as the relevant
semigroup, whose elements satisfy the following multiplication law:
\begin{equation}
\begin{tabular}{l|lll}
$\lambda _{2}$ & $\lambda _{2}$ & $\lambda _{2}$ & $\lambda _{2}$ \\
$\lambda _{1}$ & $\lambda _{2}$ & $\lambda _{1}$ & $\lambda _{2}$ \\
$\lambda _{0}$ & $\lambda _{0}$ & $\lambda _{2}$ & $\lambda _{2}$ \\ \hline
& $\lambda _{0}$ & $\lambda _{1}$ & $\lambda _{2}$%
\end{tabular}
\label{MLSL1}
\end{equation}%
with $\lambda _{2}=0_{S}$ being the zero element of the semigroup. Let $S_{L}^{(1)}=S_0\cup S_1$ be a decomposition of the semigroup%
\begin{align}
S_{0} &=\left\{ \lambda _{0},\lambda _{1},\lambda _{2}\right\} \,,  &
S_{1} &=\left\{ \lambda _{1},\lambda _{2}\right\} \,,
\end{align}%
which is said to be resonant \cite{Izaurieta:2006zz} since it satisfies the same structure as the subspaces of the $\mathfrak{osp}\left(1|4\right)$ superalgebra \eqref{SD}. The $\mathfrak{sp}\left(4\right)\otimes\mathfrak{osp}\left(1|4\right)$ superalgebra spanned by $\{T_{a}^{\pm},T_{abc}^{\pm},\mathcal{G}_{\alpha a}\}$ is obtained after considering a resonant $S_{L}^{(1)}$-expansion of the $\mathfrak{osp}\left(1|4\right)$ superalgebra and performing a $0_{S}$-reduction. The expanded generators are related to the $\mathfrak{osp}\left(1|4\right)$ ones through the semigroup elements as
\begin{equation}
    \begin{tabular}{lll}
%\cline{2-2}\cline{3-3}
\multicolumn{1}{l|}{$\lambda_2$} & \multicolumn{1}{|l}{\cellcolor[gray]{0.8}} & \multicolumn{1}{|l|}{\cellcolor[gray]{0.8}} \\ \hline
\multicolumn{1}{l|}{$\lambda_1$} & \multicolumn{1}{|l}{$T_{a}^{+}$,\ \,$T_{abc}^{+}$} & \multicolumn{1}{|l|}{$\mathcal{G}_{\alpha a}$} \\ \hline
\multicolumn{1}{l|}{$\lambda_0$} & \multicolumn{1}{|l}{$T_{a}^{-}$,\ \,$T_{abc}^{-}$} & \multicolumn{1}{|l|}{\cellcolor[gray]{0.8}} \\ \hline
\multicolumn{1}{l|}{} & \multicolumn{1}{|l}{$T_{a},\ T_{abc}$} & \multicolumn{1}{|l|}{$\mathcal{G}_{\alpha a}$}
\end{tabular}\label{EXP2}%
\end{equation}
The $\mathfrak{sp}\left(4\right)\otimes\mathfrak{osp}\left(1|4\right)$ superalgebra appears using the multiplication law of the semigroup $S_{L}^{\left(1\right)}$ \eqref{MLSL1} along the original (anti-)commutators of the $\mathfrak{osp}\left(1|4\right)$ superalgebra. In particular, $T_{a}^{-}$, $T_{abc}^{-}$ stand for the generators of the $\mathfrak{sp}\left(4\right)$ algebra, while $T_{a}^{+}$, $T_{abc}^{+}$, $\mathcal{G}_{\alpha a}$ span the $\mathfrak{osp}\left(1|4\right)$ superalgebra \eqref{osp14}. Using Theorem VII of \cite{Izaurieta:2006zz}, one can show that the $\mathfrak{sp}\left(4\right)\otimes\mathfrak{osp}\left(1|4\right)$ superalgebra admits the following bilinear invariant trace:
\begin{align}
\left\langle T^{+}_{a}T^{+}_{b}\right\rangle &=\mu\eta _{ab}\,, \notag \\ \left\langle T^{-}_{a}T^{-}_{b}\right\rangle &=\nu\eta _{ab}\,, \notag \\
\left\langle T^{+}_{abc}T^{+}_{mnk}\right\rangle &=2\mu\left(5\eta _{m\left( a\right.
}\eta _{b|n}\eta _{|\left. c\right) k}-3\eta _{\left( ab\right. }\eta
_{|\left. c\right) \left( m\right. }\eta _{\left. nk\right) }\right) \,, \notag \\ \left\langle T^{-}_{abc}T^{-}_{mnk}\right\rangle &=2\nu\left(5\eta _{m\left( a\right.
}\eta _{b|n}\eta _{|\left. c\right) k}-3\eta _{\left( ab\right. }\eta
_{|\left. c\right) \left( m\right. }\eta _{\left. nk\right) }\right) \,,\notag \\
\left\langle \mathcal{G}_{\alpha a}\mathcal{G}_{\beta b}\right\rangle &=%
\mu\left(\frac{4}{3}C_{\alpha \beta }\eta _{ab}-\frac{2}{3}\epsilon _{abc}\left(
C\Gamma ^{c}\right) _{\alpha \beta }\right)\,.  \label{IT4a}
\end{align}
The hyper-AdS algebra is obtained after considering the following redefinition of the generators,
\begin{align}
    J_{a}&=T_{a}^{+}+T_{a}^{-}\,, & P_{a}&=\frac{1}{\ell}\left(T_{a}^{+}-T_{a}^{-}\right)\,, \notag \\
    J_{abc}&=T_{abc}^{+}+T_{abc}^{-}\,, & P_{abc}&=\frac{1}{\ell}\left(T_{abc}^{+}-T_{abc}^{-}\right)\,, \notag \\
    Q_{\alpha a }&=\sqrt{\frac{2}{\ell}}\mathcal{G}_{\alpha a}\,,
\end{align}
where the $\ell$ parameter is related to the cosmological constant through $\Lambda = - \frac{1}{\ell^2}$. Indeed, the generators $\{J_{a},P_{a},J_{abc},P_{abc},Q_{\alpha a} \}$ satisfy the following non-vanishing (anti-)commutation relations:
\begin{eqnarray}
\left[ J_{a},J_{b}\right]  &=&\epsilon _{~ab}^{m}J_{m}\,, \qquad \qquad \ \ \ \, \ \left[
J_{a},P_{b}\right] =\epsilon _{~ab}^{m}P_{m}\,, \qquad \qquad \ \ \ \, \left[P_a,P_b\right]=\frac{1}{\ell^2}\epsilon_{~ab}^{m}J_{m}\,,  \notag \\
\left[ J_{a},J_{bcd}\right]  &=&3\epsilon _{~a\left( b\right. }^{m}J_{\left.
cd\right) m}\,,\qquad \ \ \left[ J_{a},P_{bcd}\right] =3\epsilon _{~a\left(
b\right. }^{m}P_{\left. cd\right) m}\,, \qquad  \ \left[ P_{a},P_{bcd}\right] =\frac{3}{\ell^2}\epsilon _{~a\left(
b\right. }^{m}J_{\left. cd\right) m}\,, \notag \\
\left[ P_{a},J_{bcd}\right]  &=&3\epsilon _{~a\left( b\right. }^{m}P_{\left.
cd\right) m}\,,\notag\\
\left[ J_{a},Q_{\alpha b}\right] &=&\frac{1}{2}\left( \Gamma _{a}\right) _{%
\text{ }\alpha }^{\beta }Q_{\beta b}+\epsilon _{abc}Q_{\beta }^{~c}\,, \qquad \qquad \left[ P_{a},Q_{\alpha b}\right] =\frac{1}{\ell}\left(\frac{1}{2}\left( \Gamma _{a}\right) _{%
\text{ }\alpha }^{\beta }Q_{\beta b}+\epsilon _{abc}Q_{\beta }^{~c}\right)\,,
 \notag\\
 \left[ J_{abc},J_{mnk}\right]  &=&-6\left( \eta _{\left( ab\right. }\epsilon
_{~\left. c\right) \left( m\right. }^{l}\eta _{\left. nk\right) }+5\epsilon
_{~\left( m\right. |\left( a\right. }^{l}\delta _{~b}^{d}\eta _{\left.
c\right) |n}\eta _{\left. k\right) d}\right) J_{l}  \notag \\
&&+2\left( 5\epsilon _{~\left( m\right. |\left( a\right. }^{l}\delta
_{~b}^{d}J_{\left. c\right) l|n}\eta _{\left. k\right) d}-\epsilon _{~\left(
m\right. |\left( a\right. }^{l}\eta _{\left. bc\right) }J_{|\left. nk\right)
l}-\epsilon _{~\left( m\right. \left( a\right. |}^{l}J_{\left. bc\right)
l}\eta _{|\left. nk\right) }\right) \,,  \notag \\
\left[ J_{abc},P_{mnk}\right]  &=&-6\left( \eta _{\left( ab\right. }\epsilon
_{~\left. c\right) \left( m\right. }^{l}\eta _{\left. nk\right) }+5\epsilon
_{~\left( m\right. |\left( a\right. }^{l}\delta _{~b}^{d}\eta _{\left.
c\right) |n}\eta _{\left. k\right) d}\right) P_{l}  \notag \\
&&+2\left( 5\epsilon _{~\left( m\right. |\left( a\right. }^{l}\delta
_{~b}^{d}P_{\left. c\right) l|n}\eta _{\left. k\right) d}-\epsilon _{~\left(
m\right. |\left( a\right. }^{l}\eta _{\left. bc\right) }P_{|\left. nk\right)
l}-\epsilon _{~\left( m\right. \left( a\right. |}^{l}P_{\left. bc\right)
l}\eta _{|\left. nk\right) }\right) \,,  \notag \\
\left[ P_{abc},P_{mnk}\right]  &=&-\frac{6}{\ell^2}\left( \eta _{\left( ab\right. }\epsilon
_{~\left. c\right) \left( m\right. }^{l}\eta _{\left. nk\right) }+5\epsilon
_{~\left( m\right. |\left( a\right. }^{l}\delta _{~b}^{d}\eta _{\left.
c\right) |n}\eta _{\left. k\right) d}\right) J_{l}  \notag \\
&&+\frac{2}{\ell^2}\left( 5\epsilon _{~\left( m\right. |\left( a\right. }^{l}\delta
_{~b}^{d}J_{\left. c\right) l|n}\eta _{\left. k\right) d}-\epsilon _{~\left(
m\right. |\left( a\right. }^{l}\eta _{\left. bc\right) }J_{|\left. nk\right)
l}-\epsilon _{~\left( m\right. \left( a\right. |}^{l}J_{\left. bc\right)
l}\eta _{|\left. nk\right) }\right) \,,  \notag \\
\left[ J_{abc},Q_{\alpha d}\right]  &=&\left( \delta _{~d}^{k}\eta _{\left(
ab\right. |}-5\eta _{d\left( a\right. }\delta _{~b|}^{k}\right) \left(
\Gamma _{|\left. c\right) }\right) _{~\alpha }^{\beta }Q_{\beta k}+\eta
_{\left( ab\right. |}\left( \Gamma _{d}\right) _{~\alpha }^{\beta }Q_{\beta
|\left. c\right) }\,,  \notag \\
\left[ P_{abc},Q_{\alpha d}\right]  &=&\frac{1}{\ell}\left[\left( \delta _{~d}^{k}\eta _{\left(
ab\right. |}-5\eta _{d\left( a\right. }\delta _{~b|}^{k}\right) \left(
\Gamma _{|\left. c\right) }\right) _{~\alpha }^{\beta }Q_{\beta k}+\eta
_{\left( ab\right. |}\left( \Gamma _{d}\right) _{~\alpha }^{\beta }Q_{\beta
|\left. c\right) }\right]\,,  \notag \\
\left\{ Q_{\alpha a},Q_{\beta b}\right\}  &=&\frac{1}{\ell}\left[\left(J_{abc}-\frac{4}{3}\eta
_{ab}J_{c}\right) \left( C\Gamma ^{c}\right) _{\alpha \beta }+\frac{5}{3}%
\epsilon _{abc}C_{\alpha \beta }J^{c}+\frac{2}{3}J_{\left( a\right. |}\left(
C\Gamma _{|\left. b\right) }\right) _{\alpha \beta }\right] \notag \\ 
&&+\left(P_{abc}-\frac{4}{3}\eta
_{ab}P_{c}\right) \left( C\Gamma ^{c}\right) _{\alpha \beta }+\frac{5}{3}%
\epsilon _{abc}C_{\alpha \beta }P^{c}+\frac{2}{3}P_{\left( a\right. |}\left(
C\Gamma _{|\left. b\right) }\right) _{\alpha \beta }\,.
 \label{HAdS}
\end{eqnarray}
Such hyper-algebra will be denoted as the $\mathfrak{hads}$ algebra. In the vanishing cosmological constant limit $\ell\rightarrow\infty$ we recover the hyper-Poincaré algebra with spin-4 generators \eqref{HP}. Interestingly, as it was noticed in \cite{Fuentealba:2015wza}, a different redefinition of the $\mathfrak{sp}\left(4\right)\otimes\mathfrak{osp}\left(1|4\right)$ generators allows us to make contact with the hyper-Poincaré algebra without spin-4 generators in the flat limit. To this end we consider the following redefinition of the generators,
\begin{align}
    J_{a}&=T_{a}^{+}+T_{a}^{-}\,, & P_{a}&=\frac{1}{\ell}\left(T_{a}^{+}-T_{a}^{-}\right)\,, \notag \\
    J_{abc}&=\frac{1}{\sqrt{\ell}}\left(T_{abc}^{+}+T_{abc}^{-}\right)\,, & P_{abc}&=\frac{1}{\sqrt{\ell}}\left(T_{abc}^{+}-T_{abc}^{-}\right)\,, \notag \\
    Q_{\alpha a }&=\sqrt{\frac{2}{\ell}}\mathcal{G}_{\alpha a}\,. \label{AREDEF}
\end{align}
Such redefinition differs subtly from the previous one and allows us, in the vanishing cosmological constant limit $\ell\rightarrow\infty$, to obtain a different hyper-Poincaré algebra:
\begin{eqnarray}
\left[ J_{a},J_{b}\right]  &=&\epsilon _{~ab}^{m}J_{m}\,, \qquad \qquad \ \ \ \, \ \left[
J_{a},P_{b}\right] =\epsilon _{~ab}^{m}P_{m}\,,   \notag \\
\left[ J_{a},J_{bcd}\right]  &=&3\epsilon _{~a\left( b\right. }^{m}J_{\left.
cd\right) m}\,,\qquad \ \ \left[ J_{a},P_{bcd}\right] =3\epsilon _{~a\left(
b\right. }^{m}P_{\left. cd\right) m}\,,  \notag \\
\left[ J_{a},Q_{\alpha b}\right] &=&\frac{1}{2}\left( \Gamma _{a}\right) _{%
\text{ }\alpha }^{\beta }Q_{\beta b}+\epsilon _{abc}Q_{\beta }^{~c}\,, 
 \notag\\
\left[ J_{abc},P_{mnk}\right]  &=&-6\left( \eta _{\left( ab\right. }\epsilon
_{~\left. c\right) \left( m\right. }^{l}\eta _{\left. nk\right) }+5\epsilon
_{~\left( m\right. |\left( a\right. }^{l}\delta _{~b}^{d}\eta _{\left.
c\right) |n}\eta _{\left. k\right) d}\right) P_{l}  \notag \\
\left\{ Q_{\alpha a},Q_{\beta b}\right\}  &=&-\frac{4}{3}\eta
_{ab}P_{c} \left( C\Gamma ^{c}\right) _{\alpha \beta }+\frac{5}{3}%
\epsilon _{abc}C_{\alpha \beta }P^{c}+\frac{2}{3}P_{\left( a\right. |}\left(
C\Gamma _{|\left. b\right) }\right) _{\alpha \beta }\,.
 \label{HP2}
\end{eqnarray}
Interestingly, the vanishing cosmological constant limit $\ell\rightarrow\infty$ allows us to get rid of the spin-4 generators. Indeed, the previous algebra contains a subalgebra spanned by $\{J_{a},P_{a},Q_{\alpha a}\}$ corresponding to the hyper-Poicaré algebra without spin-4 generators \cite{Fuentealba:2015jma}.
%one can consider the subalgebra spanned by $\{J_{a},P_{a},Q_{\alpha a}\}$ which corresponds to the hyper-Poincaré algebra being the gauge symmetry of the hypergravity theory without spin-4 gauge fields \cite{Fuentealba:2015jma}.

A gauge invariant CS action for the hyper-AdS algebra is constructed from the gauge connection one-form for the $\mathfrak{hads}$ algebra \eqref{HAdS}:
\begin{eqnarray}
A&=&e^{a}P_{a}+\omega^{a}J_{a}+e^{abc}P_{abc}+\omega^{abc}J_{abc}+\Bar{\psi}^{a}Q_{a}\,,\label{1FHADS}
\end{eqnarray}
The corresponding curvature two-form is given by
\begin{equation}
    F_{\mathfrak{hads}}=\tilde{\mathcal{T}}^{a}P_{a}+\tilde{\mathcal{R}}^{a}J_{a}+\tilde{\mathcal{T}}^{abc}P_{abc}+\tilde{\mathcal{R}}^{abc}J_{abc}+\tilde{\nabla}\Bar{\psi}^{a}Q_{a}\,, \label{2FHADS}
\end{equation}
where
\begin{eqnarray}
\tilde{\mathcal{T}}^{a}&=&de^{a}+\epsilon _{~bc}^{a}\omega^{b}e^{c}+30\epsilon^{a}_{\ bc}\omega^{bmn}e^{c}_{\ mn}-\frac{3}{2}i\Bar{\psi}_{b}\Gamma^{a}\psi^{b}\,, \notag \\ 
\tilde{\mathcal{R}}^{a}&=&d\omega^{a}+\frac{1}{2}\epsilon _{~bc}^{a}\omega^{b}\omega^{c}+15\epsilon^{a}_{\ bc}\omega^{bmn}\omega^{c}_{\ mn}+\frac{1}{2\ell^2}\epsilon^{a}_{\ bc}e^{b}e^{c}+\frac{15}{\ell^2}\epsilon^{a}_{\ bc}e^{bmn}e^{c}_{\ mn}-\frac{3}{2\ell}i\bar{\psi}_{b}\Gamma^{a}\psi^{b}\,, \notag \\
\tilde{\mathcal{T}}^{abc}&=&de^{abc}-10\epsilon^{(a}_{\ mn}\omega^{m k|b}e^{c)n}_{\ \ \ k}+3\epsilon^{(a}_{\ mn}e^{m}\omega^{n|bc)}+3\epsilon^{(a}_{\ mn}\omega^{m}e^{n|bc)}+\frac{i}{2}\Bar{\psi}^{(a}\Gamma^{|b}\psi^{c)}\,, \notag \\
\tilde{\mathcal{R}}^{abc}&=&d\omega^{abc}-5\epsilon^{(a}_{\ mn}\omega^{m k|b}\omega^{c)n}_{\ \ \ k}+3\epsilon^{(a}_{\ mn}\omega^{m}\omega^{n|bc)}-\frac{5}{\ell^2}\epsilon^{(a}_{mn}e^{mk|b}e^{c)n}_{\ \ \ k} \notag \\
&&+\frac{3}{\ell^2}\epsilon^{(a}_{mn}e^{m}e^{n|bc}+\frac{i}{2\ell}\bar{\psi}^{(a}\Gamma^{|b}\psi^{c)}\,, \notag \\
\tilde{\nabla}\psi^{a}&=&d\psi^{a}+\frac{3}{2}\omega^{b}\Gamma_{b}\psi^{a}-\omega_{b}\Gamma^{a}\psi^{b}-5\omega^{bca}\Gamma_{b}\psi_{c}+\frac{3}{2\ell}e^{b}\Gamma_{b}\psi^{a}-\frac{1}{\ell}e_b\Gamma^{a}\psi^{b}-\frac{5}{\ell}e^{bca}\Gamma_b\psi_{c}\,. \label{2FHADSb}
\end{eqnarray}
The hyper-AdS algebra \eqref{HAdS} admits the following non-vanishing components of the invariant tensor:
\begin{eqnarray}
\left\langle J_{a}J_{b}\right\rangle &=&\alpha_0\eta _{ab}\,,  \notag \\
\left\langle J_{a}P_{b}\right\rangle &=&\alpha_1\eta _{ab}\,,  \notag \\
\left\langle P_{a}P_{b}\right\rangle &=&\frac{1}{\ell^2}\alpha_0\eta _{ab}\,,  \notag \\
\left\langle J_{abc}J_{mnk}\right\rangle &=&2\alpha_0\left(5\eta _{m\left( a\right.
}\eta _{b|n}\eta _{|\left. c\right) k}-3\eta _{\left( ab\right. }\eta
_{|\left. c\right) \left( m\right. }\eta _{\left. nk\right) }\right) \,, \notag \\
\left\langle J_{abc}P_{mnk}\right\rangle &=&2\alpha_1\left(5\eta _{m\left( a\right.
}\eta _{b|n}\eta _{|\left. c\right) k}-3\eta _{\left( ab\right. }\eta
_{|\left. c\right) \left( m\right. }\eta _{\left. nk\right) }\right) \,, \notag \\
\left\langle P_{abc}P_{mnk}\right\rangle &=&\frac{2}{\ell^2}\alpha_0\left(5\eta _{m\left( a\right.
}\eta _{b|n}\eta _{|\left. c\right) k}-3\eta _{\left( ab\right. }\eta
_{|\left. c\right) \left( m\right. }\eta _{\left. nk\right) }\right) \,, \notag \\
\left\langle Q_{\alpha a}Q_{\beta b}\right\rangle &=&%
\left(\frac{\alpha_0}{\ell}+\alpha_1\right)\left(\frac{4}{3}C_{\alpha \beta }\eta _{ab}-\frac{2}{3}\epsilon _{abc}\left(
C\Gamma ^{c}\right) _{\alpha \beta }\right)\,,  \label{IT4}
\end{eqnarray}
where the coupling constants are related to the $\mathfrak{sp}\left(4\right)\otimes\mathfrak{osp}\left(1|4\right)$ ones, $\nu$ and $\mu$, as
\begin{align}
    \alpha_0&=\mu+\nu\,, & \alpha_1&=\frac{1}{\ell}\left(\mu-\nu\right)\,.
\end{align}
Then, the CS action invariant under the $\mathfrak{hads}$ algebra \eqref{HAdS} is obtained by replacing the gauge connection one-form \eqref{1FHADS} and the non-vanishing components of the invariant tensor \eqref{IT4} in the general expression of the CS action \eqref{CS}:
  \begin{eqnarray}
     I_{\mathfrak{hads}}&=&\frac{k}{4\pi}\int \alpha_{0}\mathcal{L}_{0}+ \alpha_{0} \left(\frac{1}{\ell^2}e^{a}\mathcal{T}_{a}+\frac{10}{\ell^2}e^{abc}\mathcal{T}_{abc}+\frac{2i}{\ell}\bar{\psi}_{a}\tilde{\nabla}\psi^{a}\right)+\alpha_{1}\left[2\mathcal{R}^{a}e_{a}+20\mathcal{R}^{abc}e_{abc}\right.\notag\\
     &&\left.+\frac{1}{3\ell^2}\epsilon_{abc}\left({e^{a}e^{b}e^{c}+90e^{a}e^{bmn}e^{c}_{\ mn}-100e^{amn}e^{bk}_{\ \ m}e^{c}_{\ nk}}\right)+i\bar{\psi}_{a}\tilde{\nabla}\psi^{a} \right]\,,  \label{CSHADS}
 \end{eqnarray}
where we have omitted boundary terms. Let us note that $\mathcal{T}^{a}$ and $\mathcal{T}^{abc}$ are given in \eqref{Ts} while $\mathcal{L}_{0}$ is the exotic Lagrangian coupled to spin-4 gauge fields given in \eqref{L0L1}. On the other hand, $\mathcal{R}^{a}$ and $\mathcal{R}^{abc}$ are defined in \eqref{2FHPb}. Here, one can see that the $\mathfrak{hp}_{\left(4\right)}$ CS action is recovered in the vanishing cosmological constant limit $\ell\rightarrow\infty$. In particular, $\tilde{\nabla}\psi^{a}$ reduces in the flat limit to the hyper-Poincaré covariant derivative of the fermionic gauge fields $\nabla\psi^{a}$. Let us note that, as in the usual supergravity \cite{Achucarro:1987vz,Achucarro:1989gm}, no fermionic contributions appear on the exotic sector in the vanishing cosmological constant limit $\ell\rightarrow\infty$. The field equations for the $\mathfrak{hads}$ gravity theory are given by the vanishing of the hyper-AdS curvature two-forms \eqref{2FHADS}.

Let us note that a hyper-Poincaré gravity CS action without spin-4 gauge fields can be obtained from a $\mathfrak{hads}$ CS action considering the alternative redefinition of the $\mathfrak{sp}\left(4\right)\times\mathfrak{osp}\left(1|4\right)$ generators \eqref{AREDEF} and applying a vanishing cosmological constant limit $\ell\rightarrow\infty$. In particular, one can get rid of the spin-4 generators after extracting a hyper-Poincaré subalgebra \cite{Fuentealba:2015jma} allowing us to reproduce the hypergravity action first introduced by Aragone and Deser in \cite{Aragone:1983sz}, and subsequently studied in \cite{Fuentealba:2015jma}.
%%%%%%%%%%%%%%%%%%%%%%%%%%%%%%%%%%%%%%%%%%%%%%%%%%%%%%%%%%%%%%%%%%%%%%%%%%%%%%%%%%%%%%%%%%%%%%%%%%%
\subsection{Hyper-AdS-Lorentz gravity}
A hypersymmetric version of the so-called AdS-Lorentz algebra \cite{Soroka:2004fj,Soroka:2006aj,Diaz:2012zza,Concha:2017nca,Concha:2019lhn} can be obtained from the $\mathfrak{osp}\left(1|4\right)$ superalgebra considering $S_{\mathcal{M}}^{(4)}=\{\lambda_0,\lambda_1,\lambda_2,\lambda_3,\lambda_4\}$ as the relevant semigroup. The elements of the $S_{\mathcal{M}}^{(4)}$ semigroup satisfy the following multiplication law:
\begin{equation}
\lambda _{\alpha }\lambda _{\beta }=\left\{ 
\begin{array}{lcl}
\lambda _{\alpha +\beta }\,\,\,\, & \mathrm{if}\,\,\,\,\alpha +\beta \leq
4\,, &  \\ 
\lambda _{\alpha+\beta-4}\,\,\, & \mathrm{if}\,\,\,\,\alpha +\beta >4\,. & 
\end{array}%
\right.   \label{MLSM4}
\end{equation}
Unlike $S_{E}$ and $S_{L}$, the $S_{\mathcal{M}}$ family does not contain zero element. Let us consider now a subset decomposition $S_{\mathcal{M}}^{(4)}=S_{0}\cup S_{1}$ with
\begin{align}
    S_0&=\{\lambda_0,\lambda_2,\lambda_3\}\,, & S_1&=\{\lambda_1,\lambda_3\}\,.\label{SD4}
\end{align}
which is resonant since it satifies the same algebraic structure than the $\mathfrak{osp}\left(1|4\right)$ subspaces \eqref{SD}. Then, a new hyperalgebra appears after performing a resonant $S_{\mathcal{M}}^{(4)}$-expansion of the $\mathfrak{osp}\left(1|4\right)$ superalgebra \eqref{osp14}. The expanded generators
\begin{equation}
  \lbrace J_a,P_a,Z_a,J_{abc},P_{abc},Z_{abc},Q_{\alpha a},\Sigma_{\alpha a} \rbrace \,,  
\end{equation}
are related to the $\mathfrak{osp}\left(1|4\right)$ original ones through the semigroup elements as follow
\begin{equation}
    \begin{tabular}{lll}
%\cline{2-2}\cline{3-3}
\multicolumn{1}{l|}{$\lambda_4$} & \multicolumn{1}{|l}{$Z_a,\ Z_{abc} $} & \multicolumn{1}{|l|}{\cellcolor[gray]{0.8}} \\ \hline
\multicolumn{1}{l|}{$\lambda_3$} & \multicolumn{1}{|l}{\cellcolor[gray]{0.8}} & \multicolumn{1}{|l|}{$\Sigma_{\alpha a}$} \\ \hline
\multicolumn{1}{l|}{$\lambda_2$} & \multicolumn{1}{|l}{$P_a,\ P_{abc} $} & \multicolumn{1}{|l|}{\cellcolor[gray]{0.8}} \\ \hline
\multicolumn{1}{l|}{$\lambda_1$} & \multicolumn{1}{|l}{\cellcolor[gray]{0.8}} & \multicolumn{1}{|l|}{$Q_{\alpha a}$} \\ \hline
\multicolumn{1}{l|}{$\lambda_0$} & \multicolumn{1}{|l}{$ J_{a},\ J_{abc}$} & \multicolumn{1}{|l|}{\cellcolor[gray]{0.8}} \\ \hline
\multicolumn{1}{l|}{} & \multicolumn{1}{|l}{$T_{a},\ T_{abc}$} & \multicolumn{1}{|l|}{$\mathcal{G}_{\alpha a}$}
\end{tabular}\label{EXP2b}%
\end{equation}
One can show that the expanded generators satisfy a hyper-AdS-Lorentz algebra, which we shall denote as $\mathfrak{hadsL}$ considering the multiplication law of the semigroup \eqref{MLSM4} along the $\mathfrak{osp}\left(1|4\right)$ (anti-)commutators \eqref{osp14}. Indeed, the $\mathfrak{hadsL}$ hyper-algebra satisfies the (anti-) commutation relations of the hyper-Maxwell algebra given by \eqref{HP} and \eqref{HM} along the following ones:
\begin{eqnarray}
\left[ P_{a},Z_{b}\right]  &=&\frac{1}{\ell^{2}}\epsilon _{~ab}^{m}P_{m}\,, \qquad \qquad \qquad \ \ \, \left[
Z_{a},Z_{b}\right] =\frac{1}{\ell^2}\epsilon _{~ab}^{m}Z_{m}\,,  \notag \\
\left[ P_{a},Z_{bcd}\right] &=&\frac{3}{\ell^{2}}\epsilon _{~a\left(
b\right. }^{m}P_{\left. cd\right) m}\,,  \qquad \qquad \left[ Z_{a},P_{bcd}\right]  =\frac{3}{\ell^{2}}\epsilon _{~a\left( b\right. }^{m}P_{\left.
cd\right) m}\,, \notag \\
\left[ Z_{a},Z_{bcd}\right] &=&\frac{3}{\ell^{2}}\epsilon _{~a\left(
b\right. }^{m}Z_{\left. cd\right) m}\,, \notag \\
\left[ P_{a},\Sigma _{\alpha b}\right]  &=&\frac{1}{2\ell^2}\left( \Gamma
_{a}\right) _{\text{ }\alpha }^{\beta }Q _{\beta b}+\frac{1}{\ell^2}\epsilon
_{abc}Q _{\beta }^{~c}\,,  \notag \\
\left[ Z_{a},Q_{\alpha b}\right]  &=&\frac{1}{2\ell^{2}}\left( \Gamma _{a}\right) _{%
\text{ }\alpha }^{\beta }Q _{\beta b}+\frac{1}{\ell^2}\epsilon _{abc}Q _{\beta
}^{~c}\,, \notag \\
\left[ Z_{a},\Sigma_{\alpha b}\right]  &=&\frac{1}{2\ell^2}\left( \Gamma _{a}\right) _{%
\text{ }\alpha }^{\beta }\Sigma _{\beta b}+\frac{1}{\ell^2}\epsilon _{abc}\Sigma _{\beta
}^{~c}\,, \notag \\
\left[ P_{abc},Z_{mnk}\right]  &=&-\frac{6}{\ell^2}\left( \eta _{\left( ab\right. }\epsilon
_{~\left. c\right) \left( m\right. }^{l}\eta _{\left. nk\right) }+5\epsilon
_{~\left( m\right. |\left( a\right. }^{l}\delta _{~b}^{d}\eta _{\left.
c\right) |n}\eta _{\left. k\right) d}\right) P_{l} \notag \\
&&+\frac{2}{\ell^2}\left( 5\epsilon _{~\left( m\right. |\left( a\right. }^{l}\delta
_{~b}^{d}P_{\left. c\right) l|n}\eta _{\left. k\right) d}-\epsilon _{~\left(
m\right. |\left( a\right. }^{l}\eta _{\left. bc\right) }P_{|\left. nk\right)
l}-\epsilon _{~\left( m\right. \left( a\right. |}^{l}P_{\left. bc\right)
l}\eta _{|\left. nk\right) }\right) \,,  \notag \\
\left[ Z_{abc},Z_{mnk}\right]  &=&-\frac{6}{\ell^2}\left( \eta _{\left( ab\right. }\epsilon
_{~\left. c\right) \left( m\right. }^{l}\eta _{\left. nk\right) }+5\epsilon
_{~\left( m\right. |\left( a\right. }^{l}\delta _{~b}^{d}\eta _{\left.
c\right) |n}\eta _{\left. k\right) d}\right) Z_{l}  \notag \\
&&+\frac{2}{\ell^2}\left( 5\epsilon _{~\left( m\right. |\left( a\right. }^{l}\delta
_{~b}^{d}Z_{\left. c\right) l|n}\eta _{\left. k\right) d}-\epsilon _{~\left(
m\right. |\left( a\right. }^{l}\eta _{\left. bc\right) }Z_{|\left. nk\right)
l}-\epsilon _{~\left( m\right. \left( a\right. |}^{l}Z_{\left. bc\right)
l}\eta _{|\left. nk\right) }\right) \,, \notag \\
\left[ P_{abc},\Sigma _{\alpha d}\right]  &=&\frac{1}{\ell^2}\left( \delta _{~d}^{k}\eta
_{\left( ab\right. |}-5\eta _{d\left( a\right. }\delta _{~b|}^{k}\right)
\left( \Gamma _{|\left. c\right) }\right) _{~\alpha }^{\beta }Q _{\beta
k}+\frac{1}{\ell^2}\eta _{\left( ab\right. |}\left( \Gamma _{d}\right) _{~\alpha }^{\beta
}Q _{\beta |\left. c\right) }\,,  \notag \\
\left[ Z_{abc},Q_{\alpha d}\right]  &=&\frac{1}{\ell^2}\left( \delta _{~d}^{k}\eta _{\left(
ab\right. |}-5\eta _{d\left( a\right. }\delta _{~b|}^{k}\right) \left(
\Gamma _{|\left. c\right) }\right) _{~\alpha }^{\beta }Q _{\beta
k}+\frac{1}{\ell^2}\eta _{\left( ab\right. |}\left( \Gamma _{d}\right) _{~\alpha }^{\beta
}Q _{\beta |\left. c\right) }\,, \notag \\
\left[ Z_{abc},\Sigma_{\alpha d}\right]  &=&\frac{1}{\ell^2}\left( \delta _{~d}^{k}\eta _{\left(
ab\right. |}-5\eta _{d\left( a\right. }\delta _{~b|}^{k}\right) \left(
\Gamma _{|\left. c\right) }\right) _{~\alpha }^{\beta }\Sigma _{\beta
k}+\frac{1}{\ell^2}\eta _{\left( ab\right. |}\left( \Gamma _{d}\right) _{~\alpha }^{\beta
}\Sigma _{\beta |\left. c\right) }\,, \notag \\
\left\{ \Sigma_{\alpha a},\Sigma _{\beta b}\right\}  &=&\frac{1}{\ell^2}\left[\left( P_{abc}-\frac{4}{3}%
\eta _{ab}P_{c}\right) \left( C\Gamma ^{c}\right) _{\alpha \beta }+\frac{5}{3%
}\epsilon _{abc}C_{\alpha \beta }P^{c}+\frac{2}{3}P_{\left( a\right.
|}\left( C\Gamma _{|\left. b\right) }\right) _{\alpha \beta }\right]\,, \label{HADSL}
\end{eqnarray}%
where the $\ell$ parameter appears after considering the following rescaling of the generators
\begin{align}
J_{a}&\rightarrow J_{a}\,, &J_{abc}&\rightarrow J_{abc}\,, &Q_{a\alpha}&\rightarrow \ell^{1/2}Q_{a\alpha}\,, \notag\\
P_{a}&\rightarrow \ell P_{a}\,, &P_{abc}&\rightarrow \ell P_{abc}\,, &\Sigma_{a\alpha}&\rightarrow \ell^{3/2}\Sigma_{a\alpha}\,, \notag\\
Z_{a}&\rightarrow \ell^2 J_{a}\,, &Z_{abc}&\rightarrow \ell^2 J_{abc}\,. 
\end{align}
The hyper-AdS-Lorentz algebra can also be obtained as a deformation of the hyper-Maxwell algebra \eqref{HM} and, as we shall see, offer us an alternative way to introduce a cosmological constant to hypergravity but with a second fermionic charge. Such particularity is already present in the four-dimensional AdS-Lorentz supergravity \cite{Concha:2015tla}, since this theory can be seen as an extension with cosmological constant of the minimal Maxwell supergravity \cite{deAzcarraga:2014jpa,Concha:2014tca}. Let us note that the $\mathfrak{hadsL}$ hyper-algebra obtained here can be rewritten as the $\mathfrak{osp}\left(1|4\right)\otimes\mathfrak{osp}\left(1|4\right)\otimes\mathfrak{sp}\left(4\right)$ superalgebra by considering the following redefinition of the generators,
\begin{eqnarray}
J_{a}&=&T_a+T_{a}^{+}+T_{a}^{-}\,, \qquad \qquad P_{a}=\frac{1}{\ell}\left(T_a^+-T_a^-\right)\,, \qquad \quad \, Z_{a}=\frac{1}{\ell^{2}}\left(T_a^+ + T_a^- \right)\,, \notag \\
J_{abc}&=&T_{abc}+T_{abc}^{+}+T_{abc}^{-}\,, \qquad P_{abc}=\frac{1}{\ell}\left(T_{abc}^+-T_{abc}^-\right)\,, \qquad Z_{abc}=\frac{1}{\ell^{2}}\left(T_{abc}^+ + T_{abc}^- \right)\,, \notag \\
Q_{\alpha a}&=&\frac{1}{\sqrt{\ell}}\left(\mathcal{G}_{\alpha a}^{+}-i\mathcal{G}_{\alpha a}^{-} \right)\,, \qquad \ \ \ \Sigma_{\alpha a}=\frac{1}{\sqrt{\ell^{3}}}\left(\mathcal{G}_{\alpha a}^{+}+i\mathcal{G}_{\alpha a}^{-} \right) \,, \label{redef1}
\end{eqnarray}
where $\lbrace T_a^+,T_{abc}^+,\mathcal{G}_{\alpha a}^+ \rbrace$ and $\lbrace T_a^-,T_{abc}^-,\mathcal{G}_{\alpha a}^- \rbrace$ span each one a $\mathfrak{osp}\left(1|4\right)$ superalgebra. On the other hand, $\lbrace T_a, T_{abc} \rbrace$ are $\mathfrak{sp}\left(4\right)$ generators. Then, analogously to the AdS-Lorentz algebra defined in three spacetime dimensions \cite{Concha:2018jjj}, the present hyper-algebra can be seen as the direct sum of the hyper-AdS algebra \eqref{HAdS} and the spin-4 extension of the Lorentz algebra, given by $\mathfrak{sp}\left(4\right)$. Such behavior motivates the label "hyper-AdS-Lorentz" symmetry. Nevertheless, although the $\mathfrak{osp}\left(1|4\right)\otimes\mathfrak{osp}\left(1|4\right)\otimes\mathfrak{sp}\left(4\right)$ superalgebra seems to be the natural structure to construct a CS action, it is the basis given by $\{J_a,P_a,Z_a,J_{abc},P_{abc},Z_{abc},Q_{\alpha a},\Sigma_{\alpha a}\}$  which allows us to recover the hyper-Maxwell one in the vanishing cosmological constant limit $\ell\rightarrow\infty$.

Considering the Theorem VII of \cite{Izaurieta:2006zz}, one can show that the hyper-AdS-Lorentz algebra given by \eqref{HP}, \eqref{HM} and \eqref{HADSL} admits the non-vanishing components of the invariant tensor of the $\mathfrak{hm}_{(4)}$ hyper-algebra given by \eqref{IT2} and \eqref{IT3} along with the following ones
\begin{eqnarray}
 \langle P_{a}Z_{b} \rangle&=&\frac{\alpha_{1}}{\ell^2}\eta_{ab}\,,\notag \\
 \langle Z_{a}Z_{b} \rangle&=&\frac{\alpha_{2}}{\ell^2}\eta_{ab}\,, \notag \\
 \langle P_{abc}Z_{mnk} \rangle&=&\frac{\alpha_{1}}{\ell^2}\left(10 \eta_{m(a}\eta_{b|n}\eta_{|c)k}-6\eta_{(ab|}\eta_{|c)}(m|\eta_{|nk)} \right)\,,\notag \\
 \langle Z_{abc}Z_{mnk} \rangle&=&\frac{\alpha_{2}}{\ell^2}\left(10 \eta_{m(a}\eta_{b|n}\eta_{|c)k}-6\eta_{(ab|}\eta_{|c)}(m|\eta_{|nk)} \right)\,,\notag \\
 \langle \Sigma_{\alpha a}\Sigma_{\beta b}\rangle &=&\frac{\alpha_{1}}{\ell^2}\left(\frac{4}{3}C_{\alpha \beta} \eta_{ab}-\frac{2}{3}\epsilon_{abc}\left(C\Gamma^{c}\right)_{\alpha\beta}\right)\,.\label{IT5}
\end{eqnarray}
where the $\mathfrak{hadsL}$ constants are related to the $\mathfrak{osp}\left(1|4\right)$ one through
\begin{align}
 \alpha_0&=\lambda_0\mu\,, &\alpha_1&=\lambda_2\mu\, ,&\alpha_2&=\lambda_4\mu\,.
\end{align}
The gauge connection for the $\mathfrak{hadsL}$ hyper-algebra reads
\begin{equation}
    A=e^{a}P_{a}+\omega^{a}J_{a}+k^{a}Z_{a}+e^{abc}P_{abc}+\omega^{abc}J_{abc}+k^{abc}Z_{abc}+\Bar{\psi}^{a}Q_{a}+\Bar{\xi}^{a}\Sigma_{a}\,.\label{1FHADSL}
\end{equation}
The corresponding curvature two-form is given by
\begin{equation}
    F_{\mathfrak{hadsL}}=\hat{\mathcal{T}}^{a}P_{a}+\hat{\mathcal{R}}^{a}J_{a}+\hat{\mathcal{F}}^{a}Z_{a}+\hat{\mathcal{T}}^{abc}P_{abc}+\hat{\mathcal{R}}^{abc}J_{abc}+\hat{\mathcal{F}}^{abc}Z_{abc}+\hat{\nabla}\Bar{\psi}^{a}Q_{a}+\hat{\nabla}\Bar{\xi}^{a}\Sigma_{a}\,, \label{2FHADSL}
\end{equation}
whose components can be formulated in terms of the hyper-Maxwell ones \eqref{2FHPb} and \eqref{2FHMb} as
\begin{eqnarray}
\hat{\mathcal{T}}^{a}&=&\tilde{\mathcal{T}}^{a}+\frac{1}{\ell^2}\epsilon^{a}_{\ bc}k^{b}e^{c}+\frac{30}{\ell^2}\epsilon^{a}_{\ bc}k^{bmn}e^{c}_{\ mn}-\frac{3}{2\ell^2}i\bar{\xi}_{b}\Gamma^{a}\xi^{b}\,, \notag \\
\hat{\mathcal{R}}^{a}&=&\mathcal{R}^{a}\,, \notag \\
\hat{\mathcal{F}}^{a}&=&\tilde{\mathcal{F}}^{a}+\frac{1}{\ell^2}\epsilon^{a}_{\ bc}k^{b}k^{c}+\frac{30}{\ell^2}\epsilon^{a}_{\ bc}k^{bmn}k^{c}_{\ mn}\,, \notag \\
\hat{\mathcal{T}}^{abc}&=&\tilde{\mathcal{T}}^{abc}+\frac{3}{\ell^2}\epsilon^{(a}_{\ mn}e^{m}k^{n|bc)}+\frac{3}{\ell^2}\epsilon^{(a}_{\ mn}k^{m}e^{n|bc)}-\frac{10}{\ell^2}\epsilon^{(a}_{\ mn}e^{m k|b}k^{c)n}_{\ \ \ k}+\frac{i}{\ell^2}\bar{\xi}^{(a}\Gamma^{|b}\xi^{c)}\,, \notag \\
\hat{\mathcal{R}}^{abc}&=&\mathcal{R}^{abc}\,, \notag \\
\hat{\mathcal{F}}^{abc}&=&\tilde{\mathcal{F}}^{abc}+\frac{3}{\ell^2}\epsilon^{(a}_{\ mn}k^{m}k^{n|bc)}-\frac{5}{\ell^2}\epsilon^{(a}_{\ mn}k^{m k|b}k^{c)n}_{\ \ \ k}\,, \notag \\
\hat{\nabla}\psi^{a}&=&\nabla\psi^{a}+\frac{1}{\ell^2}\left[\frac{3}{2}e^{b}\Gamma_{b}\xi^{a}-e_{b}\Gamma^{a}\xi^{b}+\frac{3}{2}k^{b}\Gamma_{b}\psi^{a}-k_{b}\Gamma^{a}\psi^{b}-5e^{bca}\Gamma_{b}\xi_{c}-5k^{bca}\Gamma_b\psi_{c}\right]\,, \notag \\
\hat{\nabla}\xi^{a}&=&\nabla\xi^{a}+\frac{1}{\ell^2}\left[\frac{3}{2}k^{b}\Gamma_{b}\xi^{a}-k_{b}\Gamma^{a}\xi^{b}-5k^{bca}\Gamma_{b}\xi_{c}\right]\, .\label{2FHADSLb}
\end{eqnarray}
Naturally, in the vanishing cosmological constant limit $\ell\rightarrow\infty$, we recover the $\mathfrak{hm}_{(4)}$ curvature two-forms. A CS hypergravity action gauge invariant under the hyper-AdS-Lorentz algebra can be constructed considering the non-vanishing components of the bilinear invariant trace given by \eqref{IT2}, \eqref{IT3} and \eqref{IT5}; and the gauge connection one-form \eqref{1FHADSL} in the CS expression \eqref{CS}. Then, the $\mathfrak{hadsL}$ CS action reduces, up to a surface term, to
\begin{eqnarray}
I_{\mathfrak{hadsL}}&=&\frac{k}{4\pi}\int\alpha_1\left[2\mathcal{R}^{a}e_{a}+20\mathcal{R}^{abc}e_{abc}+i\Bar{\psi}_{a}\hat{\nabla}\psi^{a}+\frac{1}{\ell^2}\left(2\mathcal{F}^{a}e_{a}+20\mathcal{F}^{abc}e_{abc}+i\bar{\xi}_{a}\hat{\nabla}\xi^{a}\right)\right]\,\notag\\
&&+\alpha_2\left[2\mathcal{R}^{a}k_{a}+e^{a}\mathcal{T}_{a}+20\mathcal{R}^{abc}k_{abc}+10 e^{abc}\mathcal{T}_{abc}+2i\Bar{\psi}_{a}\hat{\nabla}\xi^{a}+2i\Bar{\xi}_{a}\hat{\nabla}\psi^{a}\right.\notag\\
&&\left.+\frac{1}{\ell^{2}}\left(\mathcal{F}^{a}k_{a}+10\mathcal{F}^{abc}k_{abc}\right)\right]+\alpha_0\mathcal{L}_{0}\,,
\end{eqnarray}
where $\mathcal{L}_{0}$ is the exotic hypergravity action defined in \eqref{L0L1}. Here, $\mathcal{T}^{a}$, $\mathcal{T}^{abc}$, $\mathcal{F}^{a}$ and $\mathcal{F}^{abc}$ are the bosonic counterparts of the $\hat{\mathcal{T}}^{a}$, $\hat{\mathcal{T}}^{abc}$, $\hat{\mathcal{F}}^{a}$ and $\hat{\mathcal{F}}^{abc}$ curvatures, respectively. In particular, they are given by
\begin{eqnarray}
\mathcal{T}^{a}&=&de^{a}+\epsilon _{~bc}^{a}\omega^{b}e^{c}+30\epsilon^{a}_{\ bc}\omega^{bmn}e^{c}_{\ mn}+\frac{1}{\ell^2}\epsilon^{a}_{\ bc}k^{b}e^{c}+\frac{30}{\ell^2}\epsilon^{a}_{\ bc}k^{bmn}e^{c}_{\ mn}\,, \notag \\ 
\mathcal{T}^{abc}&=&de^{abc}-10\epsilon^{(a}_{\ mn}\omega^{m k|b}e^{c)n}_{\ \ \ k}+3\epsilon^{(a}_{\ mn}e^{m}\omega^{n|bc)}+3\epsilon^{(a}_{\ mn}\omega^{m}e^{n|bc)}+\frac{3}{\ell^2}\epsilon^{(a}_{\ mn}e^{m}k^{n|bc)}\notag\\
&&+\frac{3}{\ell^2}\epsilon^{(a}_{\ mn}k^{m}e^{n|bc)}-\frac{10}{\ell^2}\epsilon^{(a}_{\ mn}e^{m k|b}k^{c)n}_{\ \ \ k}\,, \notag \\
\mathcal{F}^{a}&=&dk^{a}+\epsilon _{~bc}^{a}\omega^{b}k^{c}+\frac{1}{2}
\epsilon _{~bc}^{a}e^{b}e^{c}+30\epsilon^{a}_{\ bc}\omega^{bmn}k^{c}_{\ mn}+15\epsilon^{a}_{\ bc}e^{bmn}e^{c}_{\ mn}+\frac{1}{\ell^2}\epsilon^{a}_{\ bc}k^{b}k^{c}\notag\\
&&+\frac{30}{\ell^2}\epsilon^{a}_{\ bc}k^{bmn}k^{c}_{\ mn}\,, \notag \\
\mathcal{F}^{abc}&=&dk^{abc}-10\epsilon^{(a}_{\ mn}\omega^{m k|b}k^{c)n}_{\ \ \ k}-5\epsilon^{(a}_{\ mn}e^{m k|b}e^{c)n}_{\ \ \ k}+3\epsilon^{(a}_{\ mn}\omega^{m}k^{n|bc)}+3\epsilon^{(a}_{\ mn}k^{m}\omega^{n|bc)} \notag \\
&&+3\epsilon^{(a}_{\ mn}e^{m}e^{n|bc)}+\frac{3}{\ell^2}\epsilon^{(a}_{\ mn}k^{m}k^{n|bc)}-\frac{5}{\ell^2}\epsilon^{(a}_{\ mn}k^{m k|b}k^{c)n}_{\ \ \ k}\,. 
\end{eqnarray}
One can see that the $\mathfrak{hm}_{(4)}$ CS gravity theory \eqref{CSHM} is recovered in the vanishing cosmological constant limit $\ell\rightarrow\infty$. In presence of a cosmological constant, the equations of motion are given by the vanishing of the $\mathfrak{hadsL}$ curvature two-forms \eqref{2FHADSLb}. Unlike the hyper-Maxwell theory studied in the previous section, the $k^{ab}$ gauge field modifies both the $\alpha_{1}$ and $\alpha_{2}$ sectors of the theory. The presence of the $k^{ab}$ gauge field in the spin-2 AdS-Lorentz CS gravity theory has not only implications at the dynamics level but also at the boundary symmetry level. Indeed, the CS gravity theory based on the AdS-Lorentz algebra possess a BTZ type black hole solution \cite{Hoseinzadeh:2014bla}, while its asymptotic symmetry is described, after considering suitable boundary conditions, by a semi-simple enlargement of $\mathfrak{bms}_{3}$ algebra \cite{Concha:2018jjj}. It would be worth to explore how the hyper-$\mathfrak{bms}_{3}$ asymptotic structure \cite{Fuentealba:2015wza} of the hyper-Poincaré gravity theory is modified in the hyper-AdS-Lorentz gravity case.
%%%%%%%%%%%%%%%%%%%%%%%%%%%%%%%%%%%%%%%%%%%%%%%%%%%%%%%%%%%%%%%%%%%%%%%%%%%%%%%%%%%%%%%%%%%%%%%%%%%
\subsection{Generalized Hyper-AdS gravity}

A generalized hyper-AdS algebra appears after applying the resonant $S$-expansion to the $\mathfrak{osp}\left(1|4\right)$ superalgebra. Indeed, let us consider $S_{\mathcal{M}}^{\left(2N\right)}=\{\lambda_0,\lambda_1,\cdots,\lambda_{2N-1},\lambda_{2N}\}$ as the relevant semigroup (with $N>1$) whose elements satisfy the following multiplication law
\begin{equation}
\lambda _{\alpha }\lambda _{\beta }=\left\{ 
\begin{array}{lcl}
\lambda _{\alpha +\beta }\,\,\,\, & \mathrm{if}\,\,\,\,\alpha +\beta \leq
2N\,, &  \\ 
\lambda _{\alpha+\beta-4\left[\frac{N+1}{2}\right]}\,\,\, & \mathrm{if}\,\,\,\,\alpha +\beta >2N\,, & 
\end{array}%
\right.   \label{MLSM2N}
\end{equation}
where $\left[ x\right]$ denotes the integer part of $x$. Let $S_{\mathcal{M}}^{\left(2N\right)}=S_0\cup S_1$ be a semigroup decomposition with
\begin{eqnarray}
S_0&=&\lbrace \lambda_{2i}, \ i=0,\ldots,N\rbrace \,,\notag\\
S_1&=&\lbrace \lambda_{2i+1}, \ i=0,\ldots,N-1\rbrace\,,
\end{eqnarray}
which is said to be resonant since it satisfies the same structure than the subspace decomposition of the original $\mathfrak{osp}\left(1|4\right)$ superalgebra \eqref{SD}. Let us note that, unlike the $S_{E}^{\left(2N\right)}$ semigroup, there is no zero element in the semigroup $S_{\mathcal{M}}^{\left(2N\right)}$. Then, after performing a resonant $S_{E}^{\left(2N\right)}$-expansion to the $\mathfrak{osp}\left(1|4\right)$ superalgebra \eqref{osp14} we find an expanded hyperalgebra. The expanded generators are related to the original $\mathfrak{osp}\left(1|4\right)$ ones through the semigroup elements as follows:
\begin{align}
    J^{(i)}_{a}&=\lambda_{2i} T_{a}\,, & J^{(i)}_{abc}&=\lambda_{2i} T_{abc}\,, & Q^{(i)}_{\alpha a}&=\lambda_{2i+1}\mathcal{G}_{\alpha a}\,. \label{EG3}
\end{align} 
The explicit expanded (anti-)commutators are obtained considering the original (anti-)commutation relations of the  $\mathfrak{osp}\left(1|4\right)$ superalgebra \eqref{osp14} along the multiplication law of the semigroup $S_{\mathcal{M}}^{\left(2N\right)}$ \eqref{MLSM2N}. The novel hyperalgebra reads
\begin{align}
\left[ J^{(i)}_{a},J^{(j)}_{b}\right] &=\epsilon _{~ab}^{m}T^{(i*j)}_{m}\,,  \notag \\
\left[ J^{(i)}_{a},J^{(j)}_{bcd}\right] &=3\epsilon _{~a\left( b\right. }^{m}J^{(i*j)}_{\left.
cd\right) m}\,,  \notag \\
\left[ J^{(i)}_{a},Q^{(j)}_{\alpha b}\right] &=\frac{1}{2}\left( \Gamma
_{a}\right) _{\ \alpha }^{\beta }Q^{(i*j)}_{\beta b}+\epsilon _{abc}%
Q^{~c\,(i*j)}_{\beta }\,,  \notag \\
\left[ J^{(i)}_{abc},J^{(j)}_{mnk}\right] &=-6\left( \eta _{\left( ab\right. }\epsilon
_{~\left. c\right) \left( m\right. }^{l}\eta _{\left. nk\right) }+5\epsilon
_{~\left( m\right. |\left( a\right. }^{l}\delta _{~b}^{d}\eta _{\left.
c\right) |n}\eta _{\left. k\right) d}\right) J^{(i*j)}_{l} \notag  \\
&+2\left( 5\epsilon _{~\left( m\right. |\left( a\right. }^{l}\delta
_{~b}^{d}J^{(i*j)}_{\left. c\right) l|n}\eta _{\left. k\right) d}-\epsilon _{~\left(
m\right. |\left( a\right. }^{l}\eta _{\left. bc\right) }J^{(i*j)}_{|\left. nk\right)
l}-\epsilon _{~\left( m\right. \left( a\right. |}^{l}J^{(i*j)}_{\left. bc\right)
l}\eta _{|\left. nk\right) }\right) \,,  \notag \\
\left[ J^{(i)}_{abc},Q^{(j)}_{\alpha d}\right] &=\left( \delta _{~d}^{k}\eta
_{\left( ab\right. |}-5\eta _{d\left( a\right. }\delta _{~b|}^{k}\right)
\left( \Gamma _{|\left. c\right) }\right) _{~\alpha }^{\beta }Q^{(i*j)}%
_{\beta k}+\eta _{\left( ab\right. |}\left( \Gamma _{d}\right) _{~\alpha
}^{\beta }Q^{(i*j)}_{\beta |\left. c\right) }\,,  \notag \\
\left\{ Q^{(i)}_{\alpha a},Q^{(j)}_{\beta b}\right\} &=\left(
J^{(i*j*1)}_{abc}-\frac{4}{3}\eta _{ab}J^{(i*j*1)}_{c}\right) \left( C\Gamma ^{c}\right)
_{\alpha \beta }+\frac{5}{3}\epsilon _{abc}C_{\alpha \beta }J^{c\,(i*j*1)}+\frac{2}{3}%
J^{(i*j*1)}_{\left( a\right. |}\left( C\Gamma _{|\left. b\right) }\right) _{\alpha
\beta }\,,  \label{GHADS}
\end{align}
where we have defined the $"*"$ product as
\begin{equation}
    i \ast j=\left\{
    \begin{array}{lcl}
    i+j\,\,\,\, & \mathrm{if}\,\,\,\,i+j \leq N\,, &
    \\
    i+j-2\left[\frac{N+1}{2}\right]\,\, & \mathrm{if}\,\,\,\,i+j > N\,. &
    \end{array}
    \right. \label{starprod}
\end{equation}
The expanded hyperalgebra, which we shall denote as $\mathfrak{ghads}^{\left(N\right)}$, can be seen as the hypersymmetric extension of the so-called generalized AdS algebra \cite{Salgado:2014qqa}\footnote{Also denoted as $\mathfrak{C}_{k}$ algebra \cite{Concha:2016hbt}.}. The $\mathfrak{ghads}^{\left(N\right)}$ hyperalgebra is characterized by $N$ fermionic charges, $2N+2$ spin-2 generators and $2N+2$ spin-4 generators. For $N=2$, we recover the $\mathfrak{hadsL}$ hyperalgebra obtained previously, after considering the rescaling of the generators
\begin{align}
J_{a}^{\left(i\right)}&\rightarrow \ell^{i}J_{a}^{\left(i\right)}\,, &J_{abc}^{\left(i\right)}&\rightarrow \ell^{i} J_{abc}^{\left(i\right)}\,, &Q_{a\alpha}^{\left(i\right)}&\rightarrow \ell^{\frac{1}{2}+i}Q_{a\alpha}^{\left(i\right)}\,. \label{resc}
\end{align}
On the other hand, $\mathfrak{ghads}^{\left(3\right)}$ corresponds to a hypersymmetric version of the $\mathfrak{C}_{5}$ algebra which has been useful to recover Pure Lovelock gravity \cite{Concha:2016kdz}. 

It is interesting to note that the $\mathfrak{ghp}^{\left(N\right)}$ hyperalgebra can be recovered from the $\mathfrak{ghads}^{\left(N\right)}$ one through an Inönü-Wigner contraction procedure \cite{Inonu:1953sp}. Indeed, the (anti-)commutators of the $\mathfrak{ghp}^{\left(N\right)}$ hyperalgebra appear after considering the rescaling of the $\mathfrak{ghads}^{\left(N\right)}$ generators as in \eqref{resc} and performing the limit $\ell\rightarrow\infty$. Diagram \eqref{diagram} summarizes the expansion and contraction relations existing between both structures:
\begin{equ}[!ht]
\begin{equation*}
\begin{tabular}{ccc}
\cline{3-3}

&  & \multicolumn{1}{|c|}{Hyper-AdS} \\ \cline{3-3}
& $\nearrow _{S_{L}^{\left( 1\right) }}$ &  \\ \cline{1-1}

\multicolumn{1}{|c}{$\mathfrak{osp}\left(1|4\right)$}& \multicolumn{1}{|c}{} & $\downarrow $ \
$\ell \rightarrow \infty $ \\ \cline{1-1}
& $\searrow ^{S_{E}^{\left( 2\right) }}$ &  \\ \cline{3-3}

&  & \multicolumn{1}{|c|}{Hyper-Poincaré} \\ \cline{3-3}
\end{tabular}%
\overset{%
\begin{array}{c}
\text{{\small Generalization ($N>1$)}}%
\end{array}%
}{\longrightarrow }%
\begin{tabular}{ccc}
\cline{3-3}

&  & \multicolumn{1}{|c|}{$\mathfrak{ghads}^{\left(N\right)}$} \\
 \cline{3-3}
& $\nearrow _{S_{\mathcal{M}}^{\left( 2N\right) }}$ &  \\ \cline{1-1}

\multicolumn{1}{|c}{$\mathfrak{osp}\left(1|4\right)$}& \multicolumn{1}{|c}{} & $\downarrow$ \ IW \\ \cline{1-1}
& $\searrow ^{S_{E}^{\left( 2N\right) }}$ &  \\ \cline{3-3}

&  & \multicolumn{1}{|c|}{$\mathfrak{ghp}^{\left(N\right)}$} \\
\cline{3-3}
\end{tabular} 
\end{equation*}
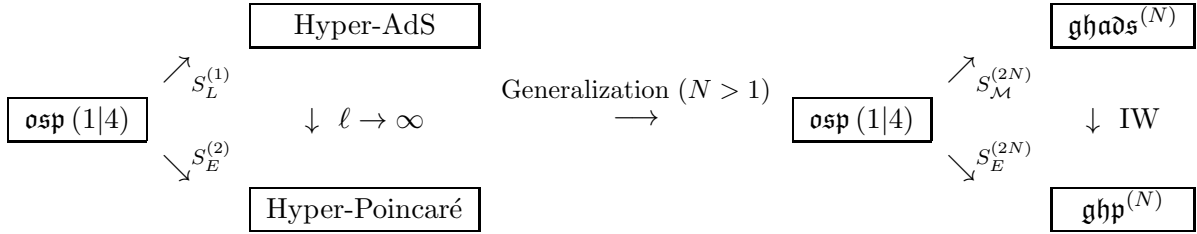
\caption{S-expansions of the $\mathfrak{osp}\left(1|4\right)$ superalgebra carried out in terms of suitable semigroups. }
\label{diagram}
\end{equ}
Similar relations also appear in the context of spin-3 algebras \cite{Caroca:2017izc}, asymptotic symmetries \cite{Caroca:2019dds} and non-relativistic symmetries \cite{Concha:2020eam}. %Let us note that the $\mathfrak{ghads}^{\left(1\right)}$ case corresponding to the hyper-AdS algebra is not obtained with the $S_{\mathcal{M}}^{(2)}$ semigroup. Such difference of the semigroup in the $N=1$ case is due to the presence of fermionic generators. Indeed in the bosonic case, the semigroup required to obtain AdS algebra from $\mathfrak{so}\left(2,1\right)$ is the $S_{\mathcal{M}}^{(1)}$ one. 

Considering Theorem VII of \cite{Izaurieta:2006zz}, one can show that the generalized hyper-AdS algebra admits the following non-vanishing components of the invariant bilinear trace:
\begin{eqnarray}
\left\langle J^{(i)}_{a}J^{(j)}_{b}\right\rangle &=&\alpha_{i*j}\eta _{ab}\,,  \notag \\
\left\langle J^{(i)}_{abc}J^{(j)}_{mnk}\right\rangle &=&2\alpha_{i*j}\left(5\eta _{m\left( a\right.
}\eta _{b|n}\eta _{|\left. c\right) k}-3\eta _{\left( ab\right. }\eta
_{|\left. c\right) \left( m\right. }\eta _{\left. nk\right) }\right) \,, \notag \\
\left\langle Q^{(i)}_{\alpha a}Q^{(j)}_{\beta b}\right\rangle &=&%
\alpha_{i*j}\left(\frac{4}{3}C_{\alpha \beta }\eta _{ab}-\frac{2}{3}\epsilon _{abc}\left(
C\Gamma ^{c}\right) _{\alpha \beta }\right)\,,  \label{ITGHADS}
\end{eqnarray}
where $\alpha_{i*j}$ is an arbitrary constant related to the $\mathfrak{osp}\left(1|4\right)$ one through the semigroup elements as
\begin{eqnarray}
\alpha_{i*j}&=&\lambda_{i*j}\mu\,,
\end{eqnarray}
whose $*$ product has been defined in \eqref{starprod}. On the other hand, the gauge connection one-form for the $\mathfrak{ghads}^{\left(N\right)}$ hyperalgebra reads
\begin{eqnarray}
A&=&\sum_{i=0}^{N}\left(\omega^{a\,(i)}J^{(i)}_{a}+\omega^{abc\,(i)}J^{(i)}_{abc}\right)+\sum_{i=0}^{N-1}\left( \bar{\psi}^{a\,(i)}Q^{(i)}_{a}\right)\,,\label{1FGHADS}
\end{eqnarray}
Then, by replacing the gauge connection one-form \eqref{1FGHADS} and the invariant tensor \eqref{ITGHADS} in the general expression of the CS form \eqref{CS}, we get the following $\mathfrak{ghads}^{\left(N\right)}$ CS hypergravity action:
\begin{eqnarray}
I_{\mathfrak{ghads}^{\left(N\right)}}&=&\frac{k}{4\pi}\int \sum_{i=0}^{N}\alpha_{i}\mathcal{L}_{i}\,,
\end{eqnarray}
where
\begin{align}
\mathcal{L}_{i}&= \omega^{(j)}_{a}d\omega^{a\,(k)}\delta_{j*k}^{i}+ \frac{1}{3}\epsilon _{~bc}^{a}\omega^{(j)}_{a}\omega^{b\,(k)}\omega^{c\,(l)}\delta_{j*k*l}^{i}+\bar{\psi}^{(j)}_{a}\nabla\psi^{a\,(k)}\delta_{j*k*1}^{i} \notag\\
&+\left(d\omega^{abc\,(j)}\delta_{j*l}^{i}-\frac{10}{3}\epsilon^{a}_{\ mn}\omega^{mpb\,(j)}\omega^{cn\,(k)}_{\ \ p}\delta_{j*k*l}^{i}+3\epsilon^{a}_{\ mn}\omega^{m\,(j)}\omega^{nbc\,(k)}\delta_{j*k*l}^{i}\right)\omega^{(l)}_{abc} \,. \label{LI2}
\end{align}
Here, the expanded covariant derivative of the spinor 1-form for the generalized hyper-AdS algebra reads
\begin{align}
    \nabla\psi^{a\,(i)}&=d\psi^{a\,(i)}+\frac{3}{2}\left(\omega^{b\,(j)}\Gamma_{b}\psi^{a\,(k)}-\omega^{(j)}_{b}\Gamma^{a}\psi^{b\,(k)}-5\omega^{bca\,(j)}\Gamma_{b}\psi^{(k)}_{c}\right)\delta_{j*k}^{i}\,.
\end{align}
Unlike the generalized hyper-Poincaré case, one can have contributions of the expanded gauge field in the exotic term $\mathcal{L}_{0}$ for particular values of $N$. Interestingly, the $\mathfrak{ghp}^{\left(N\right)}$ CS hypergravity action \eqref{LI} can be recovered from the $\mathfrak{ghads}^{\left(N\right)}$ in the limit $\ell\rightarrow\infty$ after rescaling the gauge fields as
\begin{align}
\omega_{a}^{\left(i\right)}&\rightarrow \ell^{i}\omega_{a}^{\left(i\right)}\,, &\omega_{abc}^{\left(i\right)}&\rightarrow \ell^{i} \omega_{abc}^{\left(i\right)}\,, &\psi_{a\alpha}^{\left(i\right)}&\rightarrow \ell^{\frac{1}{2}+i}\psi_{a\alpha}^{\left(i\right)}\,. \label{resc2}
\end{align}
Remarkably, the general expression for the $\mathfrak{ghads}^{\left(N\right)}$ hyperalgebra \eqref{LI2} can be directly obtained from the $\mathfrak{osp}\left(1|4\right)$ CS action \eqref{ospCS} by expressing the $\mathfrak{ghads}^{\left(N\right)}$ gauge fields in terms of the $\mathfrak{osp}\left(1|4\right)$ ones as in \eqref{gfresc} whose semigroup elements satisfy now the multiplication law of the $S_{\mathcal{M}}^{(2N)}$ semigroup \eqref{MLSM2N}. 

The dynamics of the $\mathfrak{ghads}^{\left(N\right)}$ hypergravity theory is characterized by the vanishing of the curvature two-forms:
\begin{eqnarray}
\sum_{i=0}^{N} F^{a\,\left(i\right)}&=& \sum_{i=0}^{N} \left[ d\omega^{a\,(i)}+\left(\frac{1}{2}\epsilon^{a}_{\ bc}\omega^{b\,(j)}\omega^{c\,(k)}+15\epsilon^{a}_{bc}\omega^{bmn\,(j)}\omega^{c\,(k)}_{\ mn}\right)\delta_{j*k}^{i}\right. \notag \\
&&\left.-\frac{3}{2}i\bar{\psi}_{b\,(j)}\Gamma^{a}\psi^{b\,(k)}\delta_{j*k*1}^{i}\right]\,, \notag \\
  \sum_{i=0}^{N}  F^{abc\,\left(i\right)}&=&\sum_{i=0}^{N} \left[ d\omega^{abc\,(i)}-\left(5\epsilon^{(a}_{\ mn}\omega^{m p|b\,(j)}\omega^{c)n\,(k)}_{\ \ \ p}-3\epsilon^{(a}_{\ mn}\omega^{m\,(j)}\omega^{n|bc)\,(k)}\right)\delta_{j*k}^{i}\right.\notag\\
    &&\left.+\frac{i}{2}\bar{\psi}^{(a\,(j)}\Gamma^{|b}\psi^{c)\,(k)}\delta_{j*k*1}^{i}\right]\,, \notag \\
   \sum_{i=0}^{N-1} \nabla\psi^{a\,\left(i\right)}&=& \sum_{i=0}^{N-1} \left[d\psi^{a\,(i)}+\left(\frac{3}{2}\omega^{b\,(j)}\Gamma_{b}\psi^{a\,(k)}-\omega_{b}^{(j)}\Gamma^{a}\psi^{b\,(k)}-5\omega^{bca\,(j)}\Gamma_{b}\psi_{c}^{(k)}\right)\delta_{j*k}^{i}\right]\,. \notag \\ \label{2FGHADS}
\end{eqnarray}

%%%%%%%%%%%%%%%%%%%%%%%%%%%%%%%%%%%%%%%%%%%%%%%%%%%%%%%%%%%%%%%%%%%%%%%%%%%%%%%%%%%%%%%%%%%%%%

\section{Discussion}\label{sec5}

The Lie algebra expansion methods have received a growing interest to approach diverse regimes of a gravity theory and to explore novel (super)gravity actions \cite{Izaurieta:2006aj,deAzcarraga:2007et,deAzcarraga:2012zv,Concha:2014tca,Concha:2015tla,Caroca:2018obf,Bergshoeff:2019ctr,deAzcarraga:2019mdn,Penafiel:2019czp,Gomis:2019nih,Fontanella:2020eje,Concha:2020ebl,Concha:2020eam,Kasikci:2021atn,Gomis:2022spp,Caroca:2022byi}. In this work, we have presented a consistent coupling of three-dimensional CS gravity theories with spin-$\frac{5}{2}$ gauge fields being invariant under different hyperalgebras. Two new families of hyperalgebras, which can be seen as generalizations of the hyper-Poincaré and the $\mathfrak{sp}\left(4\right)\otimes\mathfrak{osp}\left(1|4\right)$ algebra, have been obtained by applying the S-expansion method to the $\mathfrak{osp}\left(1|4\right)$ superalgebra. The S-expansion procedure also provides us with the non-vanishing components of the invariant bilinear trace form useful to construct gauge-invariant CS actions. Remarkably, the same semigroups considered to derive the generalized Poincaré and AdS algebras from the $\mathfrak{so}\left(2,1\right)$ one are used here. Such behavior has also been inherited to the spin-3 gravity models \cite{Caroca:2017izc}, asymptotic symmetries \cite{Caroca:2017onr,Caroca:2019dds} and non-relativistic gravity theories \cite{Concha:2019lhn,Penafiel:2019czp,Concha:2020eam}. Moreover, as their bosonic counterpart, both generalized hypergravity theories are related through a contraction process which in the AdS case corresponds to a vanishing cosmological constant limit.

It would be interesting to go further and explore the physical implications of the additional bosonic and fermionic content appearing in the generalized hypergravity actions as a consequence of the expansion. Regarding the boundary dynamics, one could expect that the asymptotic symmetries of the hypergravity theories presented here are described by generalizations of the hyper-$\mathfrak{bms}_{3}$, $W_{\left(2,\frac{5}{2},4\right)}$ and $W_{\left(2,4\right)}$  algebras which should be infinite-dimensional lifts of the generalized hyperalgebras. To this end, proper boundary conditions along the lines of \cite{Henneaux:2013dra,Bunster:2014mua} should be adequately performed. One might wonder if the asymptotic structures are related through a contraction procedure as their finite counterpart. Subsequently, one could analyze if hypersymmetric bounds can be obtained from the asymptotic hyperalgebras.

Another interesting aspect that deserves to be studied is the non- and ultra-relativistic limits of the hypergravity theories obtained here. It is known that, in presence of supersymmetry, the derivation of a Galilean limit for a given supergravity theory is a difficult task which can be overcome through the Lie algebra expansion method. As it was shown in \cite{Concha:2020eam}, the  non-relativistic version of different superalgebras can be obtained by expanding the supersymmetric extension of the Nappi-Witten algebra \cite{Nappi:1993ie,Figueroa-OFarrill:1999cmq} considering suitable semigroups. Following the same methodology, non-relativistic versions of the generalized spin-3 gravity theory have been introduced in \cite{Caroca:2022byi,Concha:2022muu}. Then, one could extend such procedure to a hypersymmetric extension of the Nappi-Witten and construct the respective non-Lorentzian hypergravity actions. One might then wonder whether a three-dimensional hyper-extended-Bargmann theory without spin 4 can be formulated \cite{Concha:2023wip}.

\appendix

\appendix

%%%%%%%%%%%%%%%%%%%%%%%%%%%%%%%%%%%%%%%%%%%%%%%%%%%%%%%%%%%%%%%%%%%%%%%%%%%%%%%%%%%%%%%%%%%%%

\section*{Acknowledgment}

This work was partially supported by the National Agency for Research and Development ANID - SIA grant No. SA77210097 and FONDECYT grants No.  1211077, 11220328, 1181031, 1221624, 1211226 and 11220486. P.C. and E.R. would like to thank to the Dirección de Investigación and Vicerectoría de Investigación of the Universidad Católica de la Santísima Concepción, Chile, for their constant support. J.M. has been supported by the MCI, AEI, FEDER (UE) grants PID2021-125700NB-C21 (“Gravity, Supergravity and Superstrings” (GRASS)) and IFT Centro de Excelencia Severo Ochoa CEX2020-001007-S.

%%%%%%%%%%%%%%%%%%%%%%%%%%%%%%%%%%%%%%%%%%%%%%%%%%%%%%%%%%%%%%%%%%%%%%%%%%%%%%%%%%%%%%%%%%%%%%%%%%%%

\bibliographystyle{fullsort}
 
\bibliography{Draft_vf}

\end{document}